# Suppression and enhancement of transcriptional noise by DNA looping

Jose M. G. Vilar[1,2,*] and Leonor Saiz[3,†]

[1] *Biophysics Unit (CSIC-UPV/EHU) and Department of Biochemistry and Molecular Biology, University of the Basque Country UPV/EHU, P.O. Box 644, 48080 Bilbao, Spain*

[2] *IKERBASQUE, Basque Foundation for Science, 48011 Bilbao, Spain*

[3] *Department of Biomedical Engineering, University of California, 451 E. Health Sciences Drive, Davis, CA 95616, USA*

## Abstract

DNA looping has been observed to enhance and suppress transcriptional noise but it is uncertain which of these two opposite effects is to be expected for given conditions. Here, we derive analytical expressions for the main quantifiers of transcriptional noise in terms of the molecular parameters and elucidate the role of DNA looping. Our results rationalize paradoxical experimental observations and provide the first quantitative explanation of landmark individual-cell measurements at the single molecule level on the classical *lac* operon genetic system [Choi et al., Science 322, 442-446 (2008)].

## I. INTRODUCTION

Gene expression, the process that leads to functional biomolecules from the information encoded in genes, is carried out by inherently stochastic events [1]. Very often, the underlying stochasticity is not effectively averaged out and noise, in the form of random fluctuations, propagates throughout the system [2-8]. A key regulatory step is the binding of transcription factors to DNA, which control how effectively the RNA polymerase transcribes the genes [9]. DNA looping has a widespread prominent role in this type of situations because it allows transcription factors to bind simultaneously single and multiple DNA sites and to contact the RNA polymerase from distal sites [9-11].

Typical approaches to study transcriptional noise have been based, among others, on stochastic simulations [12], linear noise approximations [13], Langevin dynamics [14], and analytical solutions of the Master equation [15]. The Fano factor, $F = \sigma^2 / \mu$, and coefficient of variation, $c_v = \sigma / \mu$, are the quantities most commonly used to characterize transcriptional noise in terms of mean, $\mu$, and standard deviation, $\sigma$, of the number of mRNA transcripts. Despite intensive research, explicit analytical expressions for these quantities, whether exact or approximate, remain scarce except for a few notable exceptions [15-17]. The presence of DNA looping is particularly challenging to deal with because it introduces additional complexity that, so far, has been possible to study only through stochastic simulations [18-20] or numerical calculations [21]. Without explicit analytical expressions, it is difficult to comprehend and fully understand how the different molecular components impact transcriptional noise and how to manipulate them to affect the system behavior. Experiments show that DNA looping can both enhance [22] and suppress [23] transcriptional noise but it is uncertain which of these two opposite effects is to be expected for given conditions.

Here, we obtain explicit analytical expressions for the Fano factor and coefficient of variation in terms of the molecular parameters and find conditions that determine whether DNA looping enhances or suppresses noise. We focus explicitly on the mode of regulation of the *lac* operon, the proverbial *E. coli* genetic system that regulates and produces the enzymes needed to metabolize lactose [24,25]. In this system, the main regulator is the *lac* repressor [26], which upon binding to the main operator prevents the RNA polymerase from transcribing the three genes used in lactose metabolism. We also consider the presence of an auxiliary operator where the repressor can bind specifically without preventing transcription (Fig. 1). Because the repressor has two DNA binding domains, it can bind two operators simultaneously by looping the intervening DNA.

## II. MATHEMATICAL DESCRIPTION

The canonical description of this type of multistate transcription systems considers that there is a set of transcriptional states $s$ and that mRNA, $m$, is produced at a rate $g_s$ for each transcription state [10]. Typically, the mRNA degradation rate $\lambda$ is independent of $s$. We consider 5 transcriptional states, which are labeled as illustrated in Fig. 1. The transcription rates $g_s$ can be expressed as the components of the vector

$$\mathbf{g} \equiv \begin{pmatrix} g_{\mathrm{on}} & g_{\mathrm{on}} & 0 & 0 & 0 \end{pmatrix}^{\mathrm{T}}, \tag{1}$$

which specifies that transcription takes place at a rate $g_{\mathrm{on}}$ only when the main operator is free.

Transitions between transcriptional states result from the binding and unbinding of the repressor. We specify the transition rates $k_{s,s'}$ from the state $s$ to the state $s'$ through the elements of the matrix

$$\mathbf{k} \equiv \begin{pmatrix} 0 & n_X k_{\mathrm{on}} & n_X k_{\mathrm{on}} & 0 & 0 \\ k_{\text{off-A}} & 0 & 0 & k_{\mathrm{loop}} & n_X k_{\mathrm{on}} \\ k_{\text{off-M}} & 0 & 0 & k_{\mathrm{loop}} & n_X k_{\mathrm{on}} \\ 0 & k_{\text{off-M}} & k_{\text{off-A}} & 0 & 0 \\ 0 & k_{\text{off-M}} & k_{\text{off-A}} & 0 & 0 \end{pmatrix}, \tag{2}$$

where $k_{on}$ is the association rate of the repressor for an operator; $k_{\text{off-M}}$ and $k_{\text{off-A}}$ are the dissociation rates of the repressor from the main and auxiliary operators, respectively; $k_{\mathrm{loop}}$ is the rate of loop formation when the repressor is bound to one operator; and $n_X$ is the average number of active repressors (see Appendix A). This description was originally developed in Ref. [18] and has been shown to accurately describe the *lac* operon under an exhaustive range of experimental conditions [27,28].

To obtain more compact expressions, we express the dissociation and the looping rates in terms of the repressor-operator association constants, $K_M$ and $K_A$, and looping local concentration, $n_L$, as $k_{\text{off-M}} = k_{\text{on}} / K_M$, $k_{\text{off-A}} = k_{\text{on}} / K_A$, and $k_{\text{loop}} = n_L k_{\text{on}}$.

The time evolution of the joint probability $P(m,s)$ of the number $m$ of mRNA molecules and the system state $s$ is governed by the Master equation

$$\begin{aligned}\frac{dP(m,s)}{dt} &= \sum_{s'}[k_{s',s}P(m,s') - k_{s,s'}P(m,s)] \\ &\quad + g_s[P(m-1,s) - P(m,s)] + \lambda[(m+1)P(m+1,s) - mP(m,s)],\end{aligned} \tag{3}$$

which takes into account the transitions between transcriptional states, mRNA production, and mRNA degradation.

## III. RESULTS

To compute the Fano factor and the coefficient of variation, we need to consider only the steady state. We proceed by rewriting the joint probability in terms of the marginal probability $P_s = \sum_m P(m,s)$ and conditional probability $P(m|s) = P(m,s)/P_s$, which we use to compute the conditional averages $\langle m \rangle_s = \sum_m mP(m|s)$ and $\langle m^2 \rangle_s = \sum_m m^2 P(m|s)$.

The steady state expression of the marginal probability $P_s$ is obtained by solving $0 = \sum_{s'}[k_{s',s}P_{s'} - k_{s,s'}P_s]$, which follows straightforwardly from Eq. (3). We write the solution using the statistical weights $Z_s$. In vector form, the marginal probabilities are given by

$$\mathbf{P} = \mathbf{Z} / \|\mathbf{Z}\|_1, \tag{4}$$

where $\|\mathbf{Z}\|_1$ is the expression of the partition function using the one norm. In the case of the *lac* operon, the statistical weights are

$$\mathbf{Z} \equiv \begin{pmatrix} 1 & n_X K_A & n_X K_M & n_X n_L K_A K_M & n_X^2 K_A K_M \end{pmatrix}^{\mathrm{T}}. \tag{5}$$

The steady-state equation for the conditional probability follows from substituting $P(m,s) = P(m|s)P_s$ in Eq. (3) with $dP(m,s)/dt = 0$. After using the detailed balance principle, $k_{s',s}P(s')/P(s) = k_{s,s'}$, and algebraic manipulations, we obtain

$$\begin{aligned} 0 = &\sum\nolimits_{s'} (k_{s,s'} - \delta_{s',s} \sum\nolimits_{s''} k_{s,s''}) P(m|s') \\ &+ g_s[P(m-1|s) - P(m|s)] + \lambda[(m+1)P(m+1|s) - mP(m|s)], \end{aligned} \tag{6}$$

where $\delta_{s',s}$ represents the Kronecker delta.

The conditional averages $\langle m \rangle_s$ and $\langle m^2 \rangle_s$ are obtained implicitly by multiplying Eq. (6) by $m$ and $m^2$, respectively, and summing over all the values of $m$, resulting in $0 = \sum_{s'} W_{s,s'} \langle m \rangle_{s'} + g_s - \lambda \langle m \rangle_s$ and $0 = \sum_{s'} W_{s,s'} \langle m(m-1) \rangle_{s'} + 2g_s \langle m \rangle_s - 2\lambda \langle m(m-1) \rangle_s$, with $W_{s,s'} = k_{s,s'} - \delta_{s',s} \sum_{s''} k_{s,s''}$. The explicit solutions are expressed in matrix form as

$$\begin{aligned} \langle \mathbf{m} \rangle &= (\lambda \mathbf{I} - \mathbf{W})^{-1} \mathbf{g}, \\ \langle \mathbf{m}^2 \rangle &= \langle \mathbf{m} \rangle + (2\lambda \mathbf{I} - \mathbf{W})^{-1} (2\mathbf{g} \circ \langle \mathbf{m} \rangle), \end{aligned} \tag{7}$$

where $\mathbf{I}$ is the identity matrix and $\langle \mathbf{m} \rangle$, $\langle \mathbf{m}^2 \rangle$, and $\mathbf{W}$ are the vectors and matrix with elements $\langle m \rangle_s$, $\langle m^2 \rangle_s$, and $W_{s,s'}$, respectively. The symbol $\circ$ represents the Hadamard product.

The mean $\mu$ and variance $\sigma^2$ are computed from the conditional averages as

$$\begin{aligned} \mu &= \langle \mathbf{m} \rangle \cdot \mathbf{P}, \\ \sigma^2 &= \langle \mathbf{m}^2 \rangle \cdot \mathbf{P} - \mu^2. \end{aligned} \tag{8}$$

The analytic expressions of $\mu$ and $\sigma^2$ in terms the molecular parameters follow straightforwardly from Eqs. (8) using Eqs. (4) and (7) with Eqs. (1), (2), and (5). The mean mRNA content is given by

$$\mu = \frac{g_{\text{on}} K_A n_X + g_{\text{on}}}{\lambda(K_A n_X (K_M (n_L + n_X) + 1) + K_M n_X + 1)}, \tag{9}$$

which has a relatively simple expression. The general expression of the variance, in contrast, is much more intricate and is given explicitly in the Supplementary Information (S.I.).

To simplify the general expressions, we considered first how adding a small looping contribution to the system changes the noise properties. Explicitly, we calculated the derivative of $F$ with respect to $n_L$ when $n_L$ is small:

$$\left.\frac{dF}{dn_L}\right|_{n_L\to 0} = \frac{g_{\text{on}} K_A K_M^2 n_X \left(\lambda K_M + k_{\text{on}} - K_M^2 k_{\text{on}} n_X^2\right)}{(K_A n_X + 1)(K_M n_X + 1)^2 (\lambda K_M + K_M k_{\text{on}} n_X + k_{\text{on}})^2}. \quad (10)$$

This expression shows that looping has a dual role. It decreases the Fano factor for

$$n_X^2 K_M^2 > 1 + \lambda K_M / k_{\text{on}} \ , \quad (11)$$

when the occupancy of the main site is sufficiently high, and increases it otherwise. The intuitive explanation is that the Fano factor has a maximum for intermediate occupancy of the main site and looping increases the occupancy towards this maximum for low occupancies or away from it in the opposite regime. In the case of $c_v$, this approach leads to an expression for $(dc_v / dn_L)_{n_L\to 0}$ that is positive for all parameter values. Explicitly, the derivative of $c_v$ with respect to $n_L$ for small $n_L$ is given by

$$\begin{aligned}\left.\frac{dc_v}{dn_L}\right|_{n_L\to 0} &= \frac{1}{2c_v}\left.\frac{dc_v^2}{dn_L}\right|_{n_L\to 0} \\ &= \frac{1}{2c_v}\frac{\lambda K_A K_M n_X}{(g_{\text{on}} K_A n_X + g_{\text{on}})(\lambda K_M + K_M k_{\text{on}} n_X + k_{\text{on}})^2} \\ &\quad \times\left(\lambda K_M^2 (\lambda + g_{\text{on}}) + K_M k_{\text{on}} (g_{\text{on}} + 2(\lambda + \lambda K_M n_X)) + (K_M k_{\text{on}} n_X + k_{\text{on}})^2\right),\end{aligned} \quad (12)$$

which is always positive. Therefore, the presence of looping always increases the coefficient of variation for small values of $n_L$.

Experiments in the *lac* operon have reported that looping decreases $F$ and increases $c_v$ [23]. In agreement with the experimental data, our results show that looping increases $c_v$ and, for the values of the parameters corresponding to the experimental conditions, Eq. (11) indicates that indeed the system is in a regime in which looping deceases $F$ (Fig. 2). We considered the exact analytic expressions (S.I.) for the experimental parameters of the *lac* operon, and found that these results, obtained for small values of $n_L$, also hold true for the actual value of the looping strength (Fig. 2).

We further confirmed our results with stochastic simulations (circles in Fig. 2) performed using the Doob-Gillespie algorithm [29,30] with transitions

$$
\begin{aligned}
&s \xrightarrow{k_{s,s'}} s',\\
&m \xrightarrow{g_s} m+1,\\
&m \xrightarrow{\lambda m} m-1.
\end{aligned}
\tag{13}
$$

Earlier numerical calculations showed that looping increases $F$ when it is plotted as a function of $\mu$ by changing the value of $n_X$ [21]. Indeed, we recapitulate this result from Eqs. (8) (Fig. 3). To provide analytical evidence for this effect, we considered a low affinity auxiliary operator and expanded in powers of $K_A$ the result of substituting $n_X$ as a function of $\mu$. Explicitly, the expression of $n_X$ in terms of the mean and the other parameters obtained from Eq. (9) is given by

$$
\begin{aligned}
n_X(\mu) = {} & \frac{g_{\text{on}}K_A - \lambda\mu K_A - \lambda\mu K_A K_M n_L - \lambda\mu K_M}{2\lambda\mu K_A K_M} \\
& + \frac{\sqrt{4\lambda\mu K_A K_M (g_{\text{on}} - \lambda\mu) + (K_A(\lambda\mu(K_M n_L + 1) - g_{\text{on}}) + \lambda\mu K_M)^2}}{2\lambda\mu K_A K_M},
\end{aligned}
\tag{14}
$$

which after substitution in the expression of the Fano factor and expanding in powers of $K_A$ leads to

$$
F[n_X(\mu)] = 1 + \frac{\lambda\mu K_M (g_{\text{on}} - \lambda\mu)}{g_{\text{on}}k_{\text{on}} + \lambda^2\mu K_M}\left(1 + \frac{g_{\text{on}}K_A^2 k_{\text{on}} n_L^2}{g_{\text{on}}k_{\text{on}} + \lambda^2\mu K_M}\right) + \text{O}(K_A^3)\,. \tag{15}
$$

Therefore, our results show that looping always increases $F[n_X(\mu)]$ when the affinity of the auxiliary site is sufficiently small. (Note that Eq. (9) implies $g_{\text{on}} - \lambda\mu > 0$.) However, this increase, in contrast to what was assumed recently [21], is not a universal property of $F$ as a function of $\mu$. Consider, for instance, that $\mu$ is adjusted by changing $K_M$ rather than $n_X$. From Eq. (9), we have

$$
K_M(\mu) = \frac{(K_A n_X + 1)(g_{\text{on}} - \lambda\mu)}{\lambda\mu n_X (K_A(n_L + n_X) + 1)} \tag{16}
$$

Expanding in powers of $K_A$ the result of substituting $K_M$ as a function of $\mu$, we obtain

$$
F[K_M(\mu)] = 1 + \frac{(g_{\text{on}} - \lambda\mu)^2}{\lambda(g_{\text{on}} - \lambda\mu) + g_{\text{on}}k_{\text{on}}n_X}\left(1 - \frac{g_{\text{on}}K_A k_{\text{on}} n_L n_X}{\lambda(g_{\text{on}} - \lambda\mu) + g_{\text{on}}k_{\text{on}}n_X}\right) + \text{O}(K_A^2)\,, \tag{17}
$$

which shows that looping always decreases $F[K_M(\mu)]$ when the affinity of the auxiliary site is sufficiently small. The full results from Eqs. (8) show that looping, indeed, decreases $F[K_M(\mu)]$ for the experimentally observed values of the parameters (Fig. 4). Note that, as functions of the mean, the coefficient of variation and Fano factor are both affected by looping essentially in the same way because they are related to each other through $c_v^2(\mu) = F(\mu)/\mu$ (Figs. 3 and 4).

Simulation results confirm that the way of adjusting the mean, either through $n_X$ or $K_M$, has profound differences in the system properties, including time courses (Fig. 5) and probability mass functions (Fig. 6).

## IV. DISCUSSION

In summary, we have provided the first analytical expressions that quantify transcriptional noise in terms of the molecular parameters in a system with DNA looping. Our results show that looping can enhance and suppress transcriptional noise in a way that matches the experimental observations. For fixed parameter values, looping decreases the Fano factor when the occupancy of the main site is sufficiently high and increases it in the opposite limit. In this case, the coefficient of variation always increases. When the mean is fixed, both the Fano factor and coefficient of variation behave in the same way, with conditions leading to an increase of noise, as when the average number of active repressors is used to adjust the mean, and to a decrease of noise, as when the main site association constant is used to adjust the mean. Overall, our results show that DNA looping provides a highly versatile mechanism to regulate both average and noisy properties of genetic systems.

## ACKNOWLEDGMENTS

This work was supported by the MINECO under grant FIS2012-38105 (J.M.G.V.) and the University of California, Davis (L.S.).

## APPENDIX A: ASSOCIATION RATES IN TERMS OF THE AVERAGE NUMBER OF ACTIVE REPRESSORS

Consider a cell with $N_X$ active repressors, $N_I$ repressors inactivated by an inducer $I$, and $N_O$ repressors bound to operator. Since the repressor's binding to inducer is much faster than its binding to the operator, the average number of active $n_X$ and inactive $n_I$ repressors are related to each other through $n_I = n_X f_I([I])$, where $f_I([I])$ is an increasing function of the inducer concentration $[I]$ with $f_I(0) = 0$. Conservation of the number of repressors leads to $N_T - N_O = (1 + f_I([I]))n_X$, where $N_T$ is the total number of repressors. The association rate of the repressors to the operator is given by $n_X^{(0)} k_{\text{on}}$, with $n_X^{(0)} = N_T / (1 + f_I([I]))$, for $N_O = 0$ and by $n_X^{(1)} k_{\text{on}} = n_X^{(0)} k_{\text{on}} (N_T - 1) / N_T$ for $N_O = 1$. Typical experimental values of $N_T$ range from 10 to 900 and considering $(N_T - 1) / N_T \approx 1$ is an excellent approximation for all the experimental setups. Therefore, we use $n_X^{(1)} = n_X^{(0)} = n_X$.

---

*j.vilar@ikerbasque.org

†lsaiz@ucdavis.edu

# Figure captions

FIG. 1 (color online). Transcriptional states of the *lac* operon (labeled with encircled numbers as used in the main text) and the possible transitions among them (arrows). DNA is represented by a wavy thick line with boxes for the main ($O_M$) and auxiliary ($O_A$) operators and the repressor is cartooned bound to the operators.

FIG. 2 (color online). Mean ($\mu$), Fano factor ($\sigma^2/\mu$), and coefficient of variation ($\sigma/\mu$) as functions of the inverse of the normalized average number of active repressors ($n_{\mathrm{WT}}/n_X$) for a system with (dashed line) and without (continuous line) looping. Here, $n_{\mathrm{WT}} = 10$ molec is the average number of wild type repressors. The values of the parameters are $k_{\mathrm{on}} = 0.0033$ molec$^{-1}s^{-1}$, $g_{\mathrm{on}} = 0.5$ molec $s^{-1}$, $\lambda = 0.0033$ $s^{-1}$, $n_L = 606$ molec (looping) or $n_L = 0$ molec (no looping), $K_M = 3.3$ molec$^{-1}$, and $K_A = 0.33$ molec$^{-1}$, which correspond to the estimated experimental values. The mean and variance were computed using the expressions in the S.I. resulting from Eqs. (8). The circles represent the results obtained from stochastic simulations [Eqs. (13)].

FIG. 3 (color online). Fano factor ($\sigma^2/\mu$) and coefficient of variation ($\sigma/\mu$) as functions of the mean ($\mu$) when the average number of active repressors ($n_X$) is changed for a system with (dashed line) and without (continuous line) looping. The values of the other parameters are the same as in Fig. 2.

FIG. 4 (color online). Fano factor ($\sigma^2/\mu$) and coefficient of variation ($\sigma/\mu$) as functions of the mean ($\mu$) when the affinity of the main site ($K_M$) is changed for a system with (dashed line) and without (continuous line) looping. The values of the other parameters are the same as in Fig. 2 with $n_X = 10$ molec.

FIG. 5 (color online). Representative time courses of the number of mRNA transcripts from stochastic simulations of the system [Eqs. (13)] without (left panels) and with (right panels) looping. The values of the parameters are the same as in Fig. 2 except for those indicated in each panel. Panels (a) and (b) correspond to the wild type case in the absence of inducer, panels (c) and (d) correspond to adjusting the mean by changing $n_X$ , and panels (e) and (f) correspond to adjusting the mean by changing $K_M$ .

FIG. 6 (color online). Probability mass functions for the number of mRNA transcripts from stochastic simulations for the same cases as in Fig. 5.

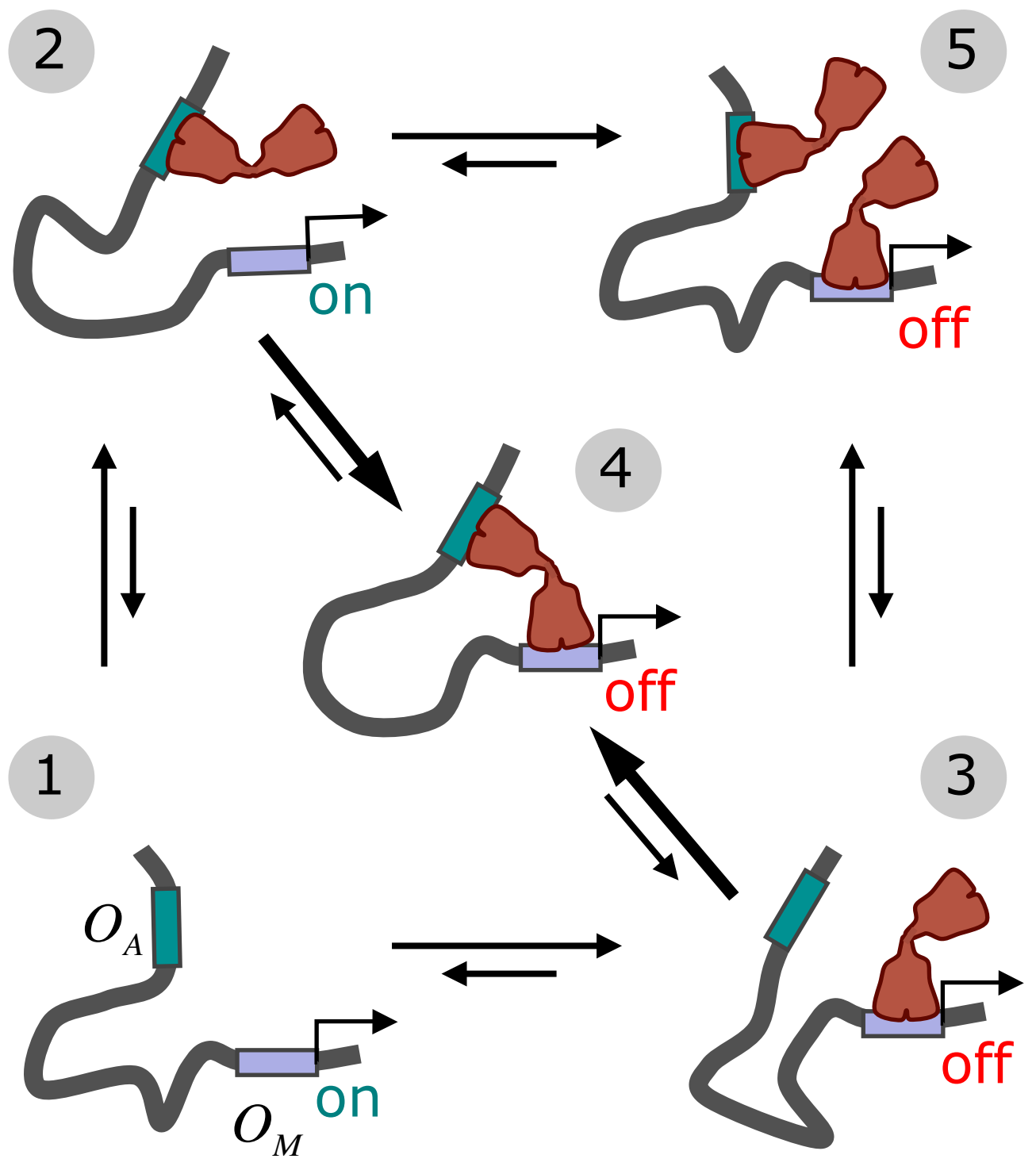

Figure 1

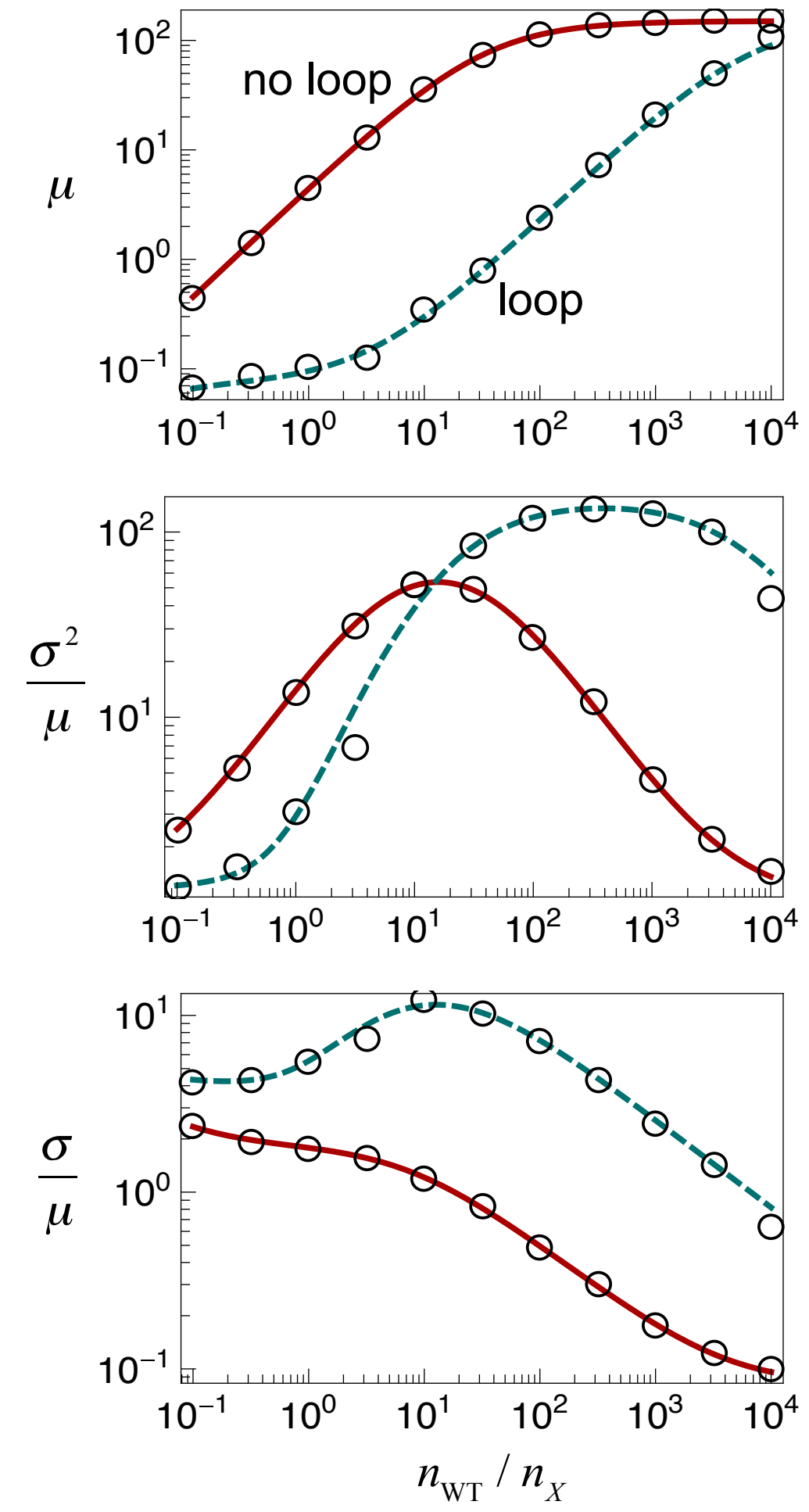

Figure 2

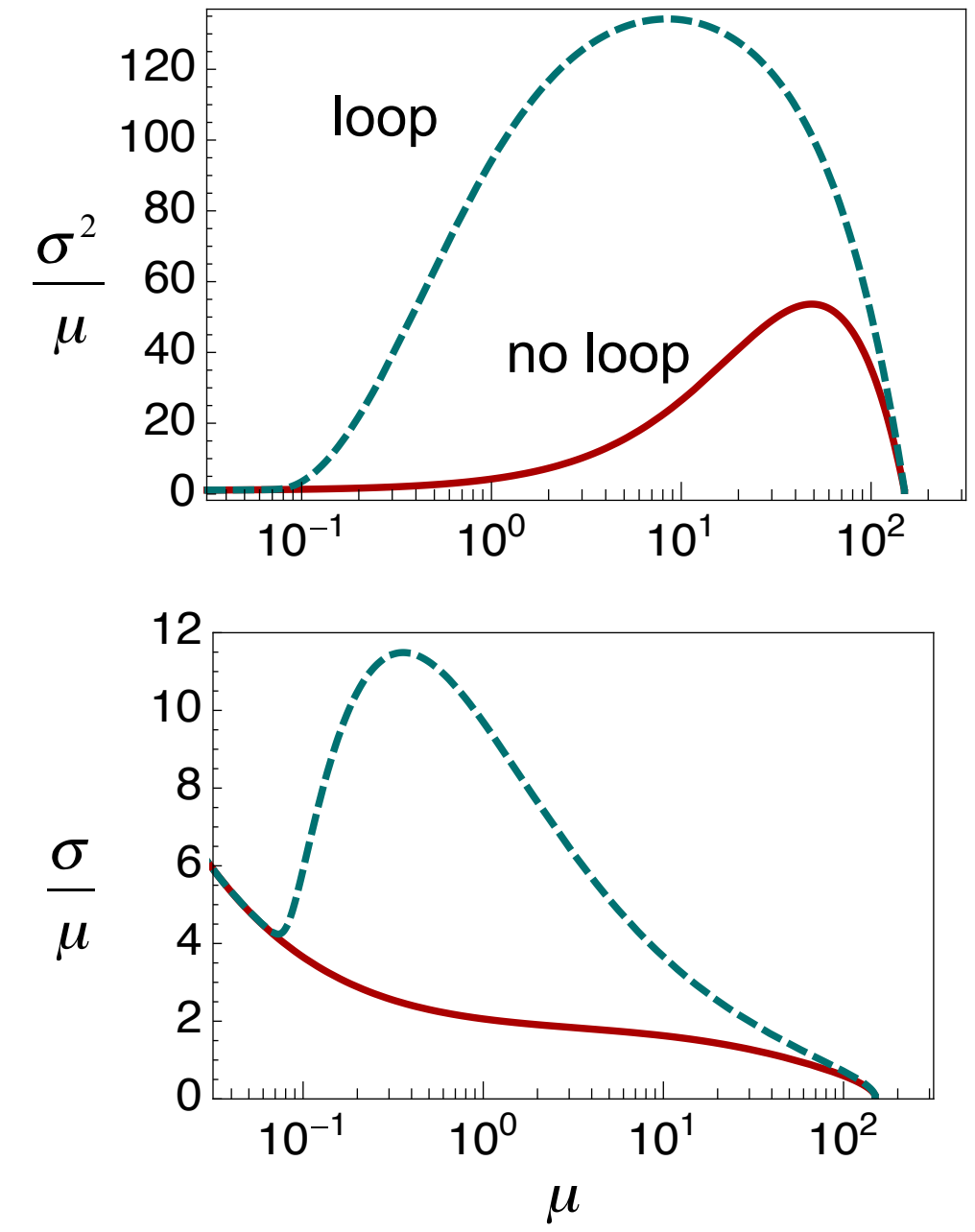

Figure 3

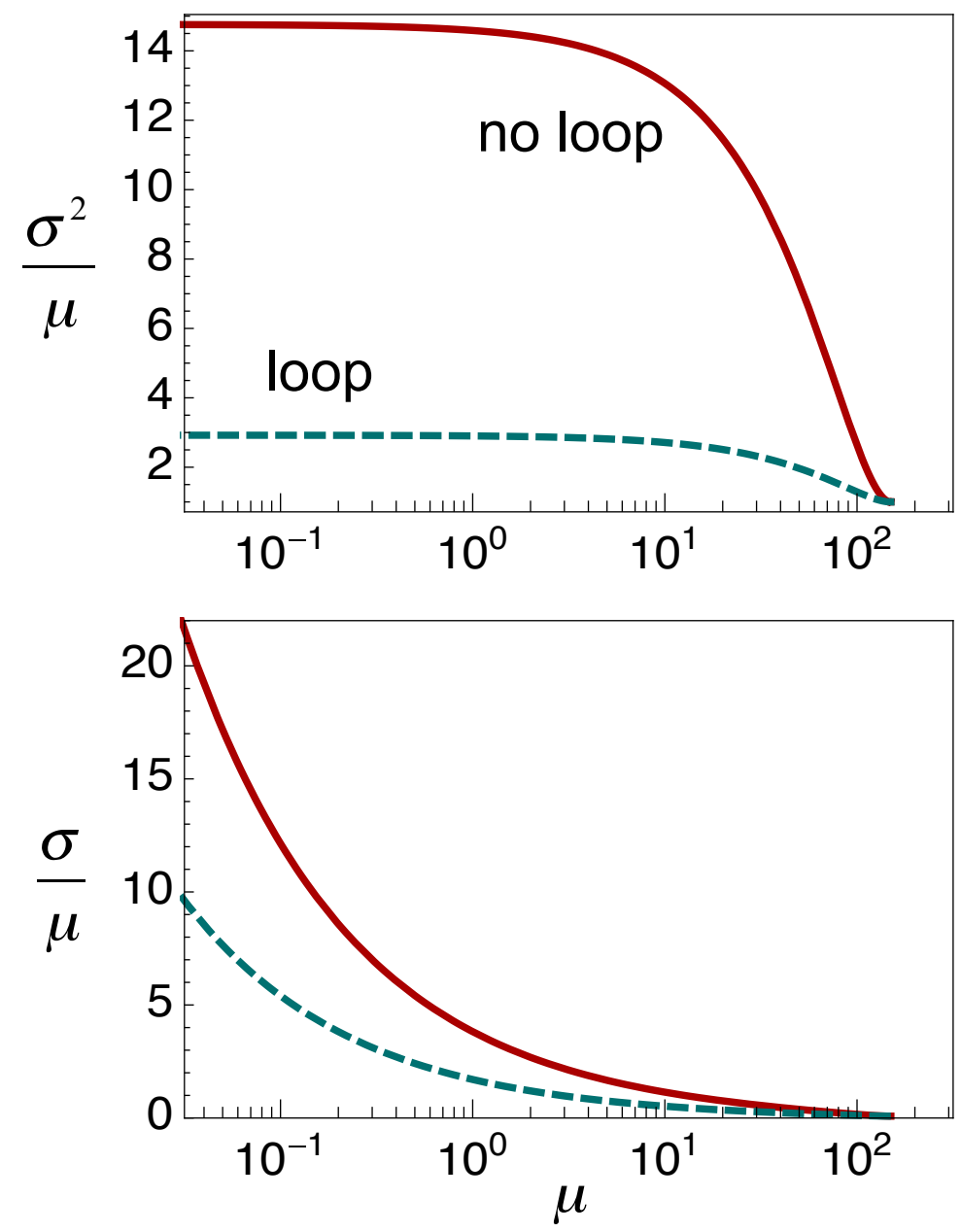

Figure 4

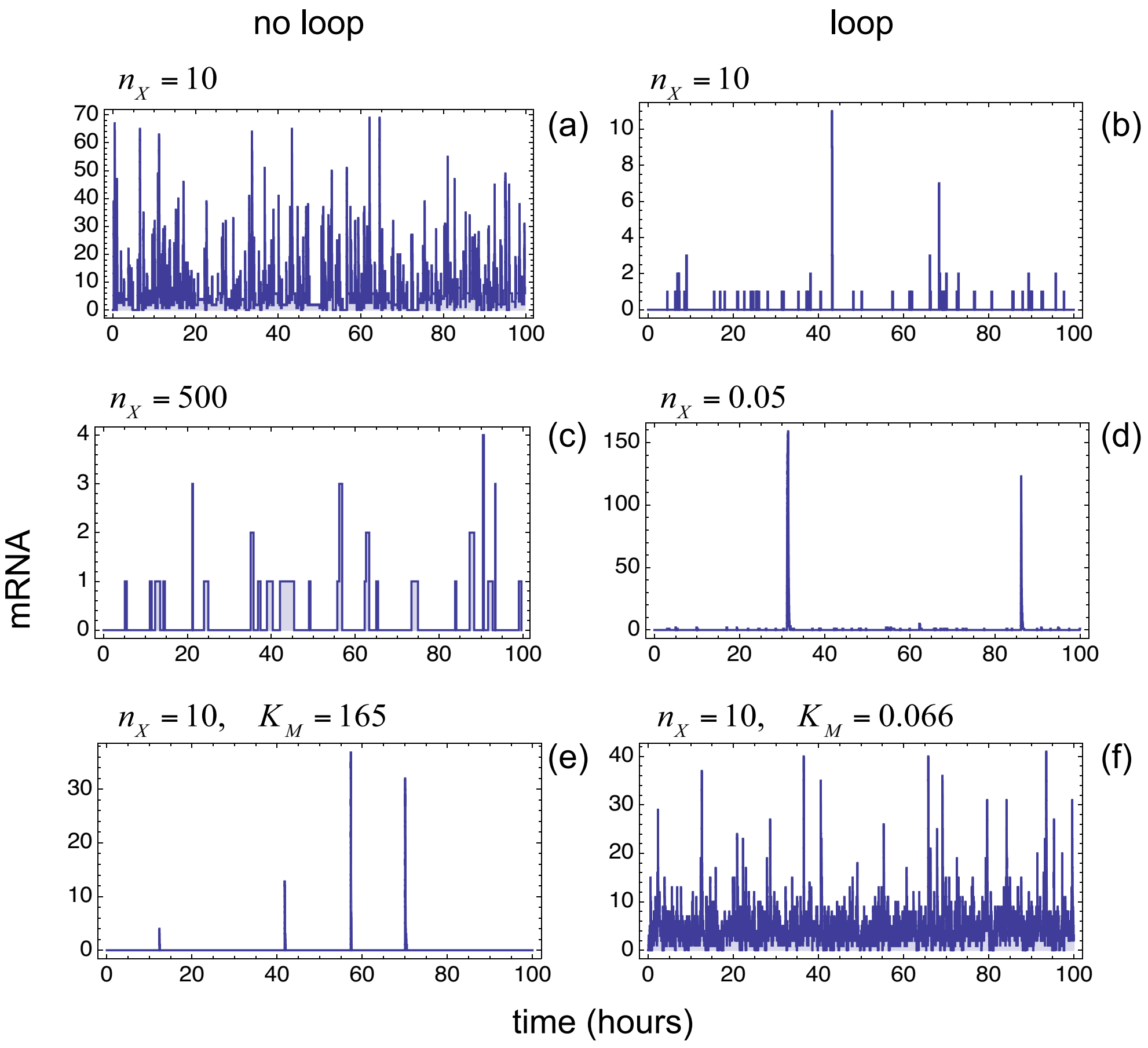

Figure 5

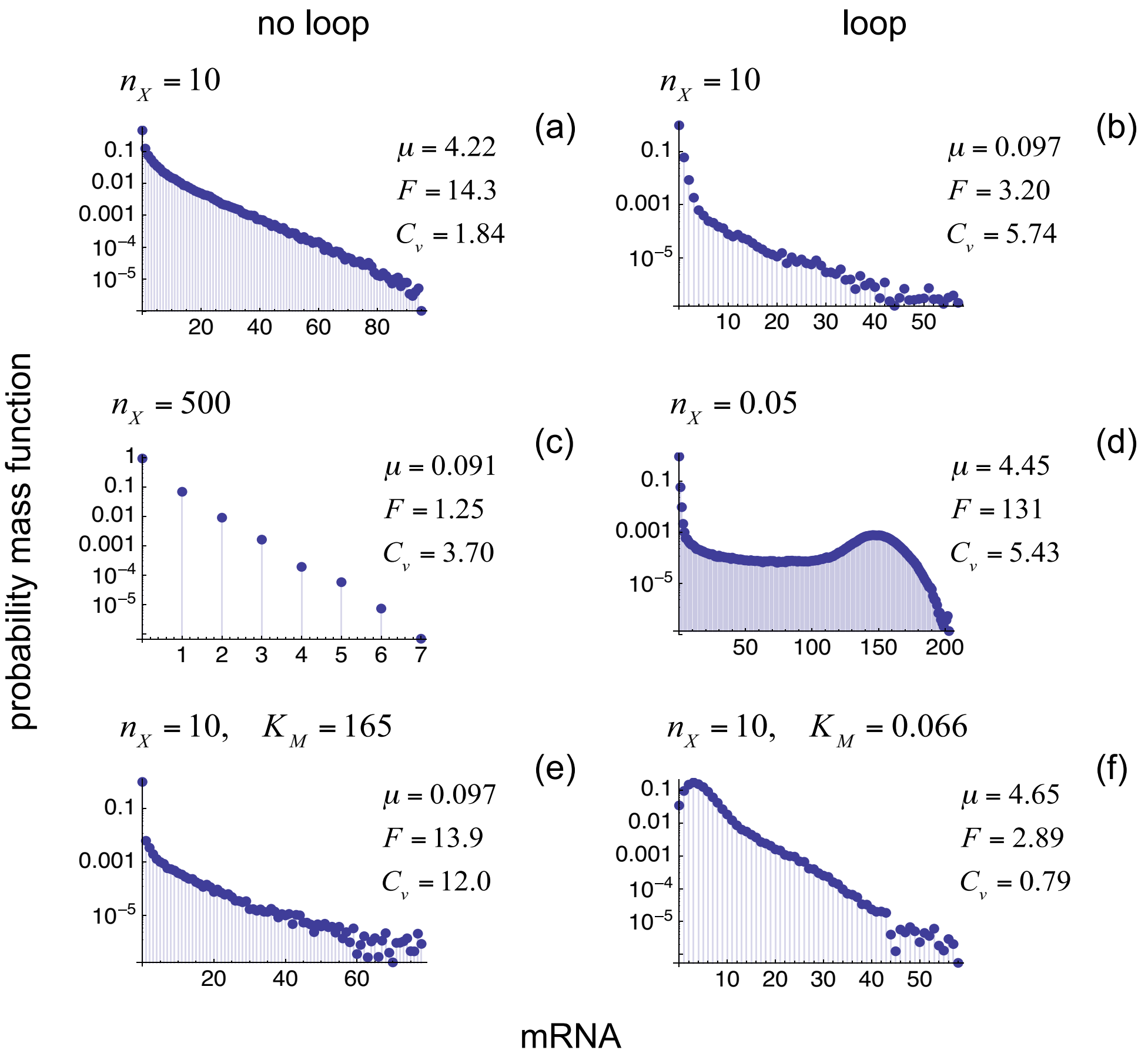

Figure 6

# Supplementary Information for "Suppression and enhancement of transcriptional noise by DNA looping"

Jose M. G. Vilar[1,2,*] and Leonor Saiz[3,†]

[1] *Biophysics Unit (CSIC-UPV/EHU) and Department of Biochemistry and Molecular Biology, University of the Basque Country UPV/EHU, P.O. Box 644, 48080 Bilbao, Spain*

[2] *IKERBASQUE, Basque Foundation for Science, 48011 Bilbao, Spain*

[3] *Department of Biomedical Engineering, University of California, 451 E. Health Sciences Drive, Davis, CA 95616, USA*

## General analytic expressions for the mean, variance, Fano factor, and coefficient of variation

The expressions for the mean and variance have been obtained directly with the computer algebra software Mathematica(R) from Eqs. (8) using Eqs. (4) and (7) with Eqs. (1), (2), and (5) in the main text. [The supplemental file "mRNAloopNoise.nb" provides the Mathematica(R) notebook with the source code.] The mean mRNA content is given by

$$\mu = \frac{g_{\text{on}} K_A n_X + g_{\text{on}}}{\lambda (K_A n_X (K_M (n_L + n_X) + 1) + K_M n_X + 1)} \ , \tag{S1}$$

which has a relatively simple expression. This expression implies $g_{\text{on}} - \lambda\mu > 0$. The general expression of the variance, in contrast, is much more intricate. It is given explicitly by

$$\sigma^2 = \frac{\sigma^2_{\text{Num}}}{\sigma^2_{\text{Den}}} \ , \tag{S2}$$

with

$$\begin{aligned}
\sigma^2_{\text{Num}} = {} & 2g_{\text{on}} K_A^4 K_M^3 k_{\text{on}}^3 n_X^6 + 7g_{\text{on}} K_A^3 K_M^3 k_{\text{on}}^3 n_X^5 + 5g_{\text{on}} K_A^4 K_M^2 k_{\text{on}}^3 n_X^5 + 2g_{\text{on}}^2 K_A^4 K_M^3 k_{\text{on}}^2 n_X^5 \\
& + 6g_{\text{on}} K_A^4 K_M^3 k_{\text{on}}^3 n_L n_X^5 + 5g_{\text{on}} K_A^4 K_M^3 k_{\text{on}}^2 \lambda n_X^5 + 9g_{\text{on}} K_A^2 K_M^3 k_{\text{on}}^3 n_X^4 + 17g_{\text{on}} K_A^3 K_M^2 k_{\text{on}}^3 n_X^4 \\
& + 4g_{\text{on}} K_A^4 K_M k_{\text{on}}^3 n_X^4 + 7g_{\text{on}}^2 K_A^3 K_M^3 k_{\text{on}}^2 n_X^4 + g_{\text{on}}^2 K_A^4 K_M^2 k_{\text{on}}^2 n_X^4 + 6g_{\text{on}} K_A^4 K_M^3 k_{\text{on}}^3 n_L^2 n_X^4 \\
& + 4g_{\text{on}} K_A^4 K_M^3 k_{\text{on}} \lambda^2 n_X^4 + 16g_{\text{on}} K_A^3 K_M^3 k_{\text{on}}^3 n_L n_X^4 + 10g_{\text{on}} K_A^4 K_M^2 k_{\text{on}}^3 n_L n_X^4 \\
& + 4g_{\text{on}}^2 K_A^4 K_M^3 k_{\text{on}}^2 n_L n_X^4 + 15g_{\text{on}} K_A^3 K_M^3 k_{\text{on}}^2 \lambda n_X^4 + 10g_{\text{on}} K_A^4 K_M^2 k_{\text{on}}^2 \lambda n_X^4 \\
& + 3g_{\text{on}}^2 K_A^4 K_M^3 k_{\text{on}} \lambda n_X^4 + 11g_{\text{on}} K_A^4 K_M^3 k_{\text{on}}^2 n_L \lambda n_X^4 + g_{\text{on}} K_A^4 k_{\text{on}}^3 n_X^3 + 5g_{\text{on}} K_A K_M^3 k_{\text{on}}^3 n_X^3 \\
& + 21g_{\text{on}} K_A^2 K_M^2 k_{\text{on}}^3 n_X^3 + 13g_{\text{on}} K_A^3 K_M k_{\text{on}}^3 n_X^3 + 2g_{\text{on}} K_A^4 K_M^3 k_{\text{on}}^3 n_L^3 n_X^3 + g_{\text{on}} K_A^4 K_M^3 \lambda^3 n_X^3 \\
& + 9g_{\text{on}}^2 K_A^2 K_M^3 k_{\text{on}}^2 n_X^3 + 3g_{\text{on}}^2 K_A^3 K_M^2 k_{\text{on}}^2 n_X^3 + 11g_{\text{on}} K_A^3 K_M^3 k_{\text{on}}^3 n_L^2 n_X^3 + 5g_{\text{on}} K_A^4 K_M^2 k_{\text{on}}^3 n_L^2 n_X^3 \\
& + 2g_{\text{on}}^2 K_A^4 K_M^3 k_{\text{on}}^2 n_L^2 n_X^3 + g_{\text{on}}^2 K_A^4 K_M^3 \lambda^2 n_X^3 + 10g_{\text{on}} K_A^3 K_M^3 k_{\text{on}} \lambda^2 n_X^3 + 6g_{\text{on}} K_A^4 K_M^2 k_{\text{on}} \lambda^2 n_X^3
\end{aligned}$$

$$
\begin{aligned}
&+6g_{\mathrm{on}}K_A^4K_M^3k_{\mathrm{on}}n_L\lambda^2n_X^3 \ +14g_{\mathrm{on}}K_A^2K_M^3k_{\mathrm{on}}^3n_Ln_X^3 \ +26g_{\mathrm{on}}K_A^3K_M^2k_{\mathrm{on}}^3n_Ln_X^3\\
&+4g_{\mathrm{on}}K_A^4K_Mk_{\mathrm{on}}^3n_Ln_X^3 \ +11g_{\mathrm{on}}^2K_A^3K_M^3k_{\mathrm{on}}^2n_Ln_X^3 \ +g_{\mathrm{on}}^2K_A^4K_M^2k_{\mathrm{on}}^2n_Ln_X^3\\
&+16g_{\mathrm{on}}K_A^2K_M^3k_{\mathrm{on}}^2\lambda n_X^3 \ +28g_{\mathrm{on}}K_A^3K_M^2k_{\mathrm{on}}^2\lambda n_X^3 \ +6g_{\mathrm{on}}K_A^4K_Mk_{\mathrm{on}}^2\lambda n_X^3\\
&+7g_{\mathrm{on}}K_A^4K_M^3k_{\mathrm{on}}^2n_L^2\lambda n_X^3 \ +8g_{\mathrm{on}}^2K_A^3K_M^3k_{\mathrm{on}}\lambda n_X^3 \ +g_{\mathrm{on}}^2K_A^4K_M^2k_{\mathrm{on}}\lambda n_X^3\\
&+24g_{\mathrm{on}}K_A^3K_M^3k_{\mathrm{on}}^2n_L\lambda n_X^3 \ +13g_{\mathrm{on}}K_A^4K_M^2k_{\mathrm{on}}^2n_L\lambda n_X^3 \ +4g_{\mathrm{on}}^2K_A^4K_M^3k_{\mathrm{on}}n_L\lambda n_X^3\\
&+3g_{\mathrm{on}}K_A^3k_{\mathrm{on}}^3n_X^2 \ +g_{\mathrm{on}}K_M^3k_{\mathrm{on}}^3n_X^2 \ +11g_{\mathrm{on}}K_AK_M^2k_{\mathrm{on}}^3n_X^2 \ +15g_{\mathrm{on}}K_A^2K_Mk_{\mathrm{on}}^3n_X^2\\
&+2g_{\mathrm{on}}K_A^3K_M^3k_{\mathrm{on}}^3n_L^3n_X^2 \ +2g_{\mathrm{on}}K_A^3K_M^3\lambda^3n_X^2 \ +g_{\mathrm{on}}K_A^4K_M^2\lambda^3n_X^2 \ +g_{\mathrm{on}}K_A^4K_M^3n_L\lambda^3n_X^2\\
&+5g_{\mathrm{on}}^2K_AK_M^3k_{\mathrm{on}}^2n_X^2 \ +3g_{\mathrm{on}}^2K_A^2K_M^2k_{\mathrm{on}}^2n_X^2 \ +5g_{\mathrm{on}}K_A^2K_M^3k_{\mathrm{on}}^3n_L^2n_X^2 \ +9g_{\mathrm{on}}K_A^3K_M^2k_{\mathrm{on}}^3n_L^2n_X^2\\
&+5g_{\mathrm{on}}^2K_A^3K_M^3k_{\mathrm{on}}^2n_L^2n_X^2 \ +2g_{\mathrm{on}}^2K_A^3K_M^3\lambda^2n_X^2 \ +2g_{\mathrm{on}}K_A^4K_M^3k_{\mathrm{on}}n_L^2\lambda^2n_X^2\\
&+8g_{\mathrm{on}}K_A^2K_M^3k_{\mathrm{on}}\lambda^2n_X^2 \ +14g_{\mathrm{on}}K_A^3K_M^2k_{\mathrm{on}}\lambda^2n_X^2 \ +2g_{\mathrm{on}}K_A^4K_Mk_{\mathrm{on}}\lambda^2n_X^2\\
&+g_{\mathrm{on}}^2K_A^4K_M^3n_L\lambda^2n_X^2 \ +10g_{\mathrm{on}}K_A^3K_M^3k_{\mathrm{on}}n_L\lambda^2n_X^2 \ +4g_{\mathrm{on}}K_A^4K_M^2k_{\mathrm{on}}n_L\lambda^2n_X^2\\
&+4g_{\mathrm{on}}K_AK_M^3k_{\mathrm{on}}^3n_Ln_X^2 \ +22g_{\mathrm{on}}K_A^2K_M^2k_{\mathrm{on}}^3n_Ln_X^2 \ +10g_{\mathrm{on}}K_A^3K_Mk_{\mathrm{on}}^3n_Ln_X^2\\
&+10g_{\mathrm{on}}^2K_A^2K_M^3k_{\mathrm{on}}^2n_Ln_X^2 \ +2g_{\mathrm{on}}^2K_A^3K_M^2k_{\mathrm{on}}^2n_Ln_X^2 \ +g_{\mathrm{on}}K_A^4K_M^3k_{\mathrm{on}}^2n_L^3\lambda n_X^2 \ +g_{\mathrm{on}}K_A^4k_{\mathrm{on}}^2\lambda n_X^2\\
&+7g_{\mathrm{on}}K_AK_M^3k_{\mathrm{on}}^2\lambda n_X^2 \ +27g_{\mathrm{on}}K_A^2K_M^2k_{\mathrm{on}}^2\lambda n_X^2 \ +15g_{\mathrm{on}}K_A^3K_Mk_{\mathrm{on}}^2\lambda n_X^2\\
&+10g_{\mathrm{on}}K_A^3K_M^3k_{\mathrm{on}}^2n_L^2\lambda n_X^2 \ +3g_{\mathrm{on}}K_A^4K_M^2k_{\mathrm{on}}^2n_L^2\lambda n_X^2 \ +g_{\mathrm{on}}^2K_A^4K_M^3k_{\mathrm{on}}n_L^2\lambda n_X^2\\
&+7g_{\mathrm{on}}^2K_A^2K_M^3k_{\mathrm{on}}\lambda n_X^2 \ +2g_{\mathrm{on}}^2K_A^3K_M^2k_{\mathrm{on}}\lambda n_X^2 \ +16g_{\mathrm{on}}K_A^2K_M^3k_{\mathrm{on}}^2n_L\lambda n_X^2\\
&+26g_{\mathrm{on}}K_A^3K_M^2k_{\mathrm{on}}^2n_L\lambda n_X^2 \ +3g_{\mathrm{on}}K_A^4K_Mk_{\mathrm{on}}^2n_L\lambda n_X^2 \ +8g_{\mathrm{on}}^2K_A^3K_M^3k_{\mathrm{on}}n_L\lambda n_X^2\\
&+g_{\mathrm{on}}^2K_A^4K_M^2k_{\mathrm{on}}n_L\lambda n_X^2 \ +3g_{\mathrm{on}}K_A^2k_{\mathrm{on}}^3n_X \ +2g_{\mathrm{on}}K_M^2k_{\mathrm{on}}^3n_X \ +7g_{\mathrm{on}}K_AK_Mk_{\mathrm{on}}^3n_X\\
&+g_{\mathrm{on}}^2K_A^3K_M^3k_{\mathrm{on}}^2n_L^3n_X \ +g_{\mathrm{on}}K_A^2K_M^3\lambda^3n_X \ +2g_{\mathrm{on}}K_A^3K_M^2\lambda^3n_X \ +g_{\mathrm{on}}K_A^3K_M^3n_L\lambda^3n_X\\
&+g_{\mathrm{on}}^2K_M^3k_{\mathrm{on}}^2n_X \ +g_{\mathrm{on}}^2K_AK_M^2k_{\mathrm{on}}^2n_X \ +4g_{\mathrm{on}}K_A^2K_M^2k_{\mathrm{on}}^3n_L^2n_X \ +3g_{\mathrm{on}}^2K_A^2K_M^3k_{\mathrm{on}}^2n_L^2n_X\\
&+g_{\mathrm{on}}^2K_A^3K_M^2k_{\mathrm{on}}^2n_L^2n_X \ +g_{\mathrm{on}}^2K_A^2K_M^3\lambda^2n_X \ +2g_{\mathrm{on}}K_A^3K_M^3k_{\mathrm{on}}n_L^2\lambda^2n_X \ +2g_{\mathrm{on}}K_AK_M^3k_{\mathrm{on}}\lambda^2n_X\\
&+10g_{\mathrm{on}}K_A^2K_M^2k_{\mathrm{on}}\lambda^2n_X \ +4g_{\mathrm{on}}K_A^3K_Mk_{\mathrm{on}}\lambda^2n_X \ +g_{\mathrm{on}}^2K_A^3K_M^3n_L\lambda^2n_X\\
&+4g_{\mathrm{on}}K_A^2K_M^3k_{\mathrm{on}}n_L\lambda^2n_X \ +6g_{\mathrm{on}}K_A^3K_M^2k_{\mathrm{on}}n_L\lambda^2n_X \ +6g_{\mathrm{on}}K_AK_M^2k_{\mathrm{on}}^3n_Ln_X\\
&+8g_{\mathrm{on}}K_A^2K_Mk_{\mathrm{on}}^3n_Ln_X \ +3g_{\mathrm{on}}^2K_AK_M^3k_{\mathrm{on}}^2n_Ln_X \ +g_{\mathrm{on}}^2K_A^2K_M^2k_{\mathrm{on}}^2n_Ln_X \ +g_{\mathrm{on}}K_A^3K_M^3k_{\mathrm{on}}^2n_L^3\lambda n_X\\
&+2g_{\mathrm{on}}K_A^3k_{\mathrm{on}}^2\lambda n_X \ +g_{\mathrm{on}}K_M^3k_{\mathrm{on}}^2\lambda n_X \ +10g_{\mathrm{on}}K_AK_M^2k_{\mathrm{on}}^2\lambda n_X \ +12g_{\mathrm{on}}K_A^2K_Mk_{\mathrm{on}}^2\lambda n_X\\
&+3g_{\mathrm{on}}K_A^2K_M^3k_{\mathrm{on}}^2n_L^2\lambda n_X \ +4g_{\mathrm{on}}K_A^3K_M^2k_{\mathrm{on}}^2n_L^2\lambda n_X \ +2g_{\mathrm{on}}^2K_A^3K_M^3k_{\mathrm{on}}n_L^2\lambda n_X\\
&+2g_{\mathrm{on}}^2K_AK_M^3k_{\mathrm{on}}\lambda n_X \ +g_{\mathrm{on}}^2K_A^2K_M^2k_{\mathrm{on}}\lambda n_X \ +3g_{\mathrm{on}}K_AK_M^3k_{\mathrm{on}}^2n_L\lambda n_X\\
&+15g_{\mathrm{on}}K_A^2K_M^2k_{\mathrm{on}}^2n_L\lambda n_X \ +5g_{\mathrm{on}}K_A^3K_Mk_{\mathrm{on}}^2n_L\lambda n_X \ +4g_{\mathrm{on}}^2K_A^2K_M^3k_{\mathrm{on}}n_L\lambda n_X\\
&+g_{\mathrm{on}}^2K_A^3K_M^2k_{\mathrm{on}}n_L\lambda n_X \ +g_{\mathrm{on}}K_Ak_{\mathrm{on}}^3 \ +g_{\mathrm{on}}K_Mk_{\mathrm{on}}^3 \ +g_{\mathrm{on}}K_A^2K_M^2\lambda^3 \ +2g_{\mathrm{on}}K_AK_M^2k_{\mathrm{on}}\lambda^2\\
&+2g_{\mathrm{on}}K_A^2K_Mk_{\mathrm{on}}\lambda^2 \ +2g_{\mathrm{on}}K_A^2K_M^2k_{\mathrm{on}}n_L\lambda^2 \ +2g_{\mathrm{on}}K_AK_Mk_{\mathrm{on}}^3n_L \ +g_{\mathrm{on}}K_A^2k_{\mathrm{on}}^2\lambda\\
&+g_{\mathrm{on}}K_M^2k_{\mathrm{on}}^2\lambda \ +3g_{\mathrm{on}}K_AK_Mk_{\mathrm{on}}^2\lambda \ +g_{\mathrm{on}}K_A^2K_M^2k_{\mathrm{on}}^2n_L^2\lambda \ +2g_{\mathrm{on}}K_AK_M^2k_{\mathrm{on}}^2n_L\lambda\\
&+2g_{\mathrm{on}}K_A^2K_Mk_{\mathrm{on}}^2n_L\lambda
\end{aligned}
$$

and

$$
\begin{aligned}
\sigma^2_{\mathrm{Den}} = {} & 2K_A^4K_M^4k_{\mathrm{on}}^3\lambda n_X^7 + 5K_A^4K_M^4k_{\mathrm{on}}^2\lambda^2n_X^6 + 7K_A^3K_M^4k_{\mathrm{on}}^3\lambda n_X^6 + 7K_A^4K_M^3k_{\mathrm{on}}^3\lambda n_X^6 \\
& + 8K_A^4K_M^4k_{\mathrm{on}}^3n_L\lambda n_X^6 + 4K_A^4K_M^4k_{\mathrm{on}}\lambda^3n_X^5 + 15K_A^3K_M^4k_{\mathrm{on}}^2\lambda^2n_X^5 + 15K_A^4K_M^3k_{\mathrm{on}}^2\lambda^2n_X^5 \\
& + 16K_A^4K_M^4k_{\mathrm{on}}^2n_L\lambda^2n_X^5 + 9K_A^2K_M^4k_{\mathrm{on}}^3\lambda n_X^5 + 24K_A^3K_M^3k_{\mathrm{on}}^3\lambda n_X^5 + 9K_A^4K_M^2k_{\mathrm{on}}^3\lambda n_X^5 \\
& + 12K_A^4K_M^4k_{\mathrm{on}}^3n_L^2\lambda n_X^5 + 21K_A^3K_M^4k_{\mathrm{on}}^3n_L\lambda n_X^5 + 21K_A^4K_M^3k_{\mathrm{on}}^3n_L\lambda n_X^5 + K_A^4K_M^4\lambda^4n_X^4 \\
& + 10K_A^3K_M^4k_{\mathrm{on}}\lambda^3n_X^4 + 10K_A^4K_M^3k_{\mathrm{on}}\lambda^3n_X^4 + 10K_A^4K_M^4k_{\mathrm{on}}n_L\lambda^3n_X^4 + 16K_A^2K_M^4k_{\mathrm{on}}^2\lambda^2n_X^4 \\
& + 43K_A^3K_M^3k_{\mathrm{on}}^2\lambda^2n_X^4 + 16K_A^4K_M^2k_{\mathrm{on}}^2\lambda^2n_X^4 + 18K_A^4K_M^4k_{\mathrm{on}}^2n_L^2\lambda^2n_X^4 + 34K_A^3K_M^4k_{\mathrm{on}}^2n_L\lambda^2n_X^4 \\
& + 34K_A^4K_M^3k_{\mathrm{on}}^2n_L\lambda^2n_X^4 + 5K_AK_M^4k_{\mathrm{on}}^3\lambda n_X^4 + 30K_A^2K_M^3k_{\mathrm{on}}^3\lambda n_X^4 + 30K_A^3K_M^2k_{\mathrm{on}}^3\lambda n_X^4 \\
& + 5K_A^4K_Mk_{\mathrm{on}}^3\lambda n_X^4 + 8K_A^4K_M^4k_{\mathrm{on}}^3n_L^3\lambda n_X^4 + 21K_A^3K_M^4k_{\mathrm{on}}^3n_L^2\lambda n_X^4 + 21K_A^4K_M^3k_{\mathrm{on}}^3n_L^2\lambda n_X^4 \\
& + 18K_A^2K_M^4k_{\mathrm{on}}^3n_L\lambda n_X^4 + 54K_A^3K_M^3k_{\mathrm{on}}^3n_L\lambda n_X^4 + 18K_A^4K_M^2k_{\mathrm{on}}^3n_L\lambda n_X^4 + 2K_A^3K_M^4\lambda^4n_X^3 \\
& + 2K_A^4K_M^3\lambda^4n_X^3 + 2K_A^4K_M^4n_L\lambda^4n_X^3 + 8K_A^4K_M^4k_{\mathrm{on}}n_L^2\lambda^3n_X^3 + 8K_A^2K_M^4k_{\mathrm{on}}\lambda^3n_X^3 \\
& + 24K_A^3K_M^3k_{\mathrm{on}}\lambda^3n_X^3 + 8K_A^4K_M^2k_{\mathrm{on}}\lambda^3n_X^3 + 16K_A^3K_M^4k_{\mathrm{on}}n_L\lambda^3n_X^3 + 16K_A^4K_M^3k_{\mathrm{on}}n_L\lambda^3n_X^3 \\
& + 8K_A^4K_M^4k_{\mathrm{on}}^2n_L^3\lambda^2n_X^3 + 7K_AK_M^4k_{\mathrm{on}}^2\lambda^2n_X^3 + 43K_A^2K_M^3k_{\mathrm{on}}^2\lambda^2n_X^3 + 43K_A^3K_M^2k_{\mathrm{on}}^2\lambda^2n_X^3 \\
& + 7K_A^4K_Mk_{\mathrm{on}}^2\lambda^2n_X^3 + 23K_A^3K_M^4k_{\mathrm{on}}^2n_L^2\lambda^2n_X^3 + 23K_A^4K_M^3k_{\mathrm{on}}^2n_L^2\lambda^2n_X^3 \\
& + 22K_A^2K_M^4k_{\mathrm{on}}^2n_L\lambda^2n_X^3 + 68K_A^3K_M^3k_{\mathrm{on}}^2n_L\lambda^2n_X^3 + 22K_A^4K_M^2k_{\mathrm{on}}^2n_L\lambda^2n_X^3 \\
& + 2K_A^4K_M^4k_{\mathrm{on}}^3n_L^4\lambda n_X^3 + K_A^4k_{\mathrm{on}}^3\lambda n_X^3 + K_M^4k_{\mathrm{on}}^3\lambda n_X^3 + 16K_AK_M^3k_{\mathrm{on}}^3\lambda n_X^3 + 36K_A^2K_M^2k_{\mathrm{on}}^3\lambda n_X^3 \\
& + 16K_A^3K_Mk_{\mathrm{on}}^3\lambda n_X^3 + 7K_A^3K_M^4k_{\mathrm{on}}^3n_L^3\lambda n_X^3 + 7K_A^4K_M^3k_{\mathrm{on}}^3n_L^3\lambda n_X^3 + 9K_A^2K_M^4k_{\mathrm{on}}^3n_L^2\lambda n_X^3 \\
& + 36K_A^3K_M^3k_{\mathrm{on}}^3n_L^2\lambda n_X^3 + 9K_A^4K_M^2k_{\mathrm{on}}^3n_L^2\lambda n_X^3 + 5K_AK_M^4k_{\mathrm{on}}^3n_L\lambda n_X^3 + 45K_A^2K_M^3k_{\mathrm{on}}^3n_L\lambda n_X^3 \\
& + 45K_A^3K_M^2k_{\mathrm{on}}^3n_L\lambda n_X^3 + 5K_A^4K_Mk_{\mathrm{on}}^3n_L\lambda n_X^3 + K_A^2K_M^4\lambda^4n_X^2 + 4K_A^3K_M^3\lambda^4n_X^2 \\
& + K_A^4K_M^2\lambda^4n_X^2 + K_A^4K_M^4n_L^2\lambda^4n_X^2 + 2K_A^3K_M^4n_L\lambda^4n_X^2 + 2K_A^4K_M^3n_L\lambda^4n_X^2 \\
& + 2K_A^4K_M^4k_{\mathrm{on}}n_L^3\lambda^3n_X^2 + 6K_A^3K_M^4k_{\mathrm{on}}n_L^2\lambda^3n_X^2 + 6K_A^4K_M^3k_{\mathrm{on}}n_L^2\lambda^3n_X^2 + 2K_AK_M^4k_{\mathrm{on}}\lambda^3n_X^2 \\
& + 18K_A^2K_M^3k_{\mathrm{on}}\lambda^3n_X^2 + 18K_A^3K_M^2k_{\mathrm{on}}\lambda^3n_X^2 + 2K_A^4K_Mk_{\mathrm{on}}\lambda^3n_X^2 + 6K_A^2K_M^4k_{\mathrm{on}}n_L\lambda^3n_X^2 \\
& + 24K_A^3K_M^3k_{\mathrm{on}}n_L\lambda^3n_X^2 + 6K_A^4K_M^2k_{\mathrm{on}}n_L\lambda^3n_X^2 + K_A^4K_M^4k_{\mathrm{on}}^2n_L^4\lambda^2n_X^2 + 4K_A^3K_M^4k_{\mathrm{on}}^2n_L^3\lambda^2n_X^2 \\
& + 4K_A^4K_M^3k_{\mathrm{on}}^2n_L^3\lambda^2n_X^2 + K_A^4k_{\mathrm{on}}^2\lambda^2n_X^2 + K_M^4k_{\mathrm{on}}^2\lambda^2n_X^2 + 17K_AK_M^3k_{\mathrm{on}}^2\lambda^2n_X^2 \\
& + 39K_A^2K_M^2k_{\mathrm{on}}^2\lambda^2n_X^2 + 17K_A^3K_Mk_{\mathrm{on}}^2\lambda^2n_X^2 + 6K_A^2K_M^4k_{\mathrm{on}}^2n_L^2\lambda^2n_X^2 + 27K_A^3K_M^3k_{\mathrm{on}}^2n_L^2\lambda^2n_X^2 \\
& + 6K_A^4K_M^2k_{\mathrm{on}}^2n_L^2\lambda^2n_X^2 + 4K_AK_M^4k_{\mathrm{on}}^2n_L\lambda^2n_X^2 + 40K_A^2K_M^3k_{\mathrm{on}}^2n_L\lambda^2n_X^2 \\
& + 40K_A^3K_M^2k_{\mathrm{on}}^2n_L\lambda^2n_X^2 + 4K_A^4K_Mk_{\mathrm{on}}^2n_L\lambda^2n_X^2 + 3K_A^3k_{\mathrm{on}}^3\lambda n_X^2 + 3K_M^3k_{\mathrm{on}}^3\lambda n_X^2 \\
& + 18K_AK_M^2k_{\mathrm{on}}^3\lambda n_X^2 + 18K_A^2K_Mk_{\mathrm{on}}^3\lambda n_X^2 + 6K_A^3K_M^3k_{\mathrm{on}}^3n_L^3\lambda n_X^2 + 15K_A^2K_M^3k_{\mathrm{on}}^3n_L^2\lambda n_X^2 \\
& + 15K_A^3K_M^2k_{\mathrm{on}}^3n_L^2\lambda n_X^2 + 12K_AK_M^3k_{\mathrm{on}}^3n_L\lambda n_X^2 + 36K_A^2K_M^2k_{\mathrm{on}}^3n_L\lambda n_X^2 + 12K_A^3K_Mk_{\mathrm{on}}^3n_L\lambda n_X^2 \\
& + 2K_A^2K_M^3\lambda^4n_X + 2K_A^3K_M^2\lambda^4n_X + 2K_A^3K_M^3n_L\lambda^4n_X + 4K_A^3K_M^3k_{\mathrm{on}}n_L^2\lambda^3n_X \\
& + 4K_AK_M^3k_{\mathrm{on}}\lambda^3n_X + 12K_A^2K_M^2k_{\mathrm{on}}\lambda^3n_X + 4K_A^3K_Mk_{\mathrm{on}}\lambda^3n_X + 8K_A^2K_M^3k_{\mathrm{on}}n_L\lambda^3n_X \\
& + 8K_A^3K_M^2k_{\mathrm{on}}n_L\lambda^3n_X + 2K_A^3K_M^3k_{\mathrm{on}}^2n_L^3\lambda^2n_X + 2K_A^3k_{\mathrm{on}}^2\lambda^2n_X + 2K_M^3k_{\mathrm{on}}^2\lambda^2n_X \\
& + 13K_AK_M^2k_{\mathrm{on}}^2\lambda^2n_X + 13K_A^2K_Mk_{\mathrm{on}}^2\lambda^2n_X + 6K_A^2K_M^3k_{\mathrm{on}}^2n_L^2\lambda^2n_X + 6K_A^3K_M^2k_{\mathrm{on}}^2n_L^2\lambda^2n_X \\
& + 6K_AK_M^3k_{\mathrm{on}}^2n_L\lambda^2n_X + 20K_A^2K_M^2k_{\mathrm{on}}^2n_L\lambda^2n_X + 6K_A^3K_Mk_{\mathrm{on}}^2n_L\lambda^2n_X + 3K_A^2k_{\mathrm{on}}^3\lambda n_X \\
& + 3K_M^2k_{\mathrm{on}}^3\lambda n_X + 8K_AK_Mk_{\mathrm{on}}^3\lambda n_X + 6K_A^2K_M^2k_{\mathrm{on}}^3n_L^2\lambda n_X + 9K_AK_M^2k_{\mathrm{on}}^3n_L\lambda n_X
\end{aligned}
$$

$$+9K_A^2K_M k_{\text{on}}^3 n_L \lambda n_X \; + K_A^2K_M^2\lambda^4 \; + 2K_A K_M^2 k_{\text{on}}\lambda^3 \; + 2K_A^2K_M k_{\text{on}}\lambda^3 \; + 2K_A^2K_M^2 k_{\text{on}} n_L\lambda^3$$
$$+K_A^2 k_{\text{on}}^2\lambda^2 \; + K_M^2 k_{\text{on}}^2\lambda^2 \; + 3K_A K_M k_{\text{on}}^2\lambda^2 \; + K_A^2K_M^2 k_{\text{on}}^2 n_L^2\lambda^2 \; + 2K_A K_M^2 k_{\text{on}}^2 n_L\lambda^2$$
$$+2K_A^2K_M k_{\text{on}}^2 n_L\lambda^2 \; + K_A k_{\text{on}}^3\lambda \; + K_M k_{\text{on}}^3\lambda \; + 2K_A K_M k_{\text{on}}^3 n_L\lambda \, .$$

The Fano factor, $F = \sigma^2 / \mu$, and coefficient of variation, $c_v = \sigma / \mu$, follow directly from the previous expressions of the mean, $\mu$, and standard deviation, $\sigma$.

# Supplementary Information for "Suppression and enhancement of transcriptional noise by DNA looping": Mathematica(R) notebook

Jose M. G. Vilar [1,2] and Leonor Saiz [3]

[1] *Biophysics Unit (CSIC-UPV/EHU) and Department of Biochemistry and Molecular Biology, University of the Basque Country UPV/EHU, P.O. Box 644, 48080 Bilbao, Spain*

[2] *IKERBASQUE, Basque Foundation for Science, 48011 Bilbao, Spain*

[3] *Department of Biomedical Engineering, University of California, 451 E. Health Sciences Drive, Davis, CA 95616, USA*

```
$Assumptions = n > 0 && Ka > 0 && Km > 0 && nl > 0 && gon > 0 && koffa > 0 &&
  koffm > 0 && kloop > 0 && λ > 0 && kon > 0 && gon > M λ && M > 0 && x > 0 && α > 0

n > 0 && Ka > 0 && Km > 0 && nl > 0 && gon > 0 && koffa > 0 &&
 koffm > 0 && kloop > 0 && λ > 0 && kon > 0 && gon > M λ && M > 0 && x > 0 && α > 0

parex := {Ka → kon / koffa, Km → kon / koffm, nl → kloop / kon, n → 1, x → koffm / koffa}
pars := {gon → 0.5, koffa → 0.01, koffm → 0.001, kloop → 2, λ → 1 / 300, kon → 0.033}
substex := {Ka → "K_A", Km → "K_M", n → "n_X", nl → "n_L", kon → "k_on", gon → "g_on", M → "μ"}

t := {gon, gon, 0, 0, 0}

       (   0          n kon       n kon       0         0    )
       ( kon / Ka     0           0           nl kon    n kon )
k :=   ( kon / Km     0           0           nl kon    n kon )
       (   0          kon / Km    kon / Ka    0         0    )
       (   0          kon / Km    kon / Ka    0         0    )

i := IdentityMatrix[Length[t]]
u := ConstantArray[1, {Length[t], Length[t]}]
kkp = (k - i (k.u))¬;

pv = (# / Total[#] &@{1, Ka n, Km n, Ka Km n nl, Ka Km n^2}) // FullSimplify;

kk := ((k - i (k.u)) KroneckerProduct[pv, 1 / pv])¬

mv = Inverse[λ i - kk].t;
m2v = mv + Inverse[2 λ i - kk].(2 t mv);

mvs = mv.pv // FullSimplify;
m2vs = m2v.pv ;

var = m2vs - mvs^2 // Simplify;
fano = (var / mvs) // Simplify;
CV2 = (var / mvs^2) // Simplify;

Print["μ=", mvs /. substex, "\n"]
Print["σ^2=", var /. substex, "\n"]
```

$$\mu=\frac{g_{on}+g_{on}K_A n_X}{(1+K_M n_X+K_A n_X(1+K_M(n_L+n_X)))\lambda}$$

$$\begin{aligned}
\sigma^2=\Big(&g_{on}\Big(K_M k_{on}^2\Big(k_{on}(1+K_M n_X)^2+K_M(g_{on}K_M n_X+\lambda+K_M n_X\lambda)\Big)+\\
&K_A^4 n_X^2\Big(k_{on}^2 n_X(1+2K_M(n_L+n_X))+k_{on}(1+K_M(n_L+3n_X))\lambda+K_M\lambda^2\Big)\\
&\quad\Big(k_{on}(1+K_M(n_L+n_X))^2+K_M(g_{on}K_M(n_L+n_X)+(1+K_M(n_L+n_X))\lambda)\Big)+\\
&K_A k_{on}\Big(k_{on}^2(1+K_M n_X)\Big(1+2K_M(n_L+3n_X)+K_M^2 n_X(4n_L+5n_X)\Big)+2K_M^2\lambda(g_{on}K_M n_X+\lambda+K_M n_X\lambda)+\\
&\quad K_M k_{on}\Big(g_{on}K_M n_X(1+3K_M n_L+5K_M n_X)+\Big(3+2K_M(n_L+5n_X)+K_M^2 n_X(3n_L+7n_X)\Big)\lambda\Big)\Big)+\\
&K_A^2\Big(k_{on}^3 n_X\Big(3+K_M(8n_L+15n_X)+K_M^3 n_X\Big(5n_L^2+14n_L n_X+9n_X^2\Big)+K_M^2\Big(4n_L^2+22n_L n_X+21n_X^2\Big)\Big)+\\
&\quad K_M^2\lambda^2(g_{on}K_M n_X+\lambda+K_M n_X\lambda)+\\
&\quad K_M k_{on}\lambda\Big(g_{on}K_M n_X(1+4K_M n_L+7K_M n_X)+2\Big(1+2K_M^2 n_X(n_L+2n_X)+K_M(n_L+5n_X)\Big)\lambda\Big)+\\
&\quad k_{on}^2\Big(g_{on}K_M^2 n_X\Big(n_L+3K_M n_L^2+3n_X+10K_M n_L n_X+9K_M n_X^2\Big)+\\
&\qquad\Big(1+2K_M(n_L+6n_X)+K_M^3 n_X\Big(3n_L^2+16n_L n_X+16n_X^2\Big)+K_M^2\Big(n_L^2+15n_L n_X+27n_X^2\Big)\Big)\lambda\Big)\Big)+\\
&K_A^3 n_X\Big(k_{on}^3 n_X\Big(3+K_M^3(n_L+n_X)^2(2n_L+7n_X)+K_M(10n_L+13n_X)+K_M^2\Big(9n_L^2+26n_L n_X+17n_X^2\Big)\Big)+\\
&\quad K_M^2\lambda^2(g_{on}K_M(n_L+2n_X)+(2+K_M n_L+2K_M n_X)\lambda)+K_M k_{on}\lambda\Big(g_{on}K_M(n_L+2n_X)\\
&\qquad(1+2K_M(n_L+2n_X))+2\Big(2+K_M(3n_L+7n_X)+K_M^2\Big(n_L^2+5n_L n_X+5n_X^2\Big)\Big)\lambda\Big)+\\
&\quad k_{on}^2\Big(g_{on}K_M^2\Big(n_L^2(1+K_M n_L)+n_L(2+5K_M n_L)n_X+(3+11K_M n_L)n_X^2+7K_M n_X^3\Big)+\\
&\qquad\Big(2+5K_M(n_L+3n_X)+K_M^2\Big(4n_L^2+26n_L n_X+28n_X^2\Big)+K_M^3\\
&\qquad\quad\Big(n_L^3+10n_L^2 n_X+24n_L n_X^2+15n_X^3\Big)\Big)\lambda\Big)\Big)\Big)\Big)\Big/\\
\Big(&(1+K_M n_X+K_A n_X(1+K_M(n_L+n_X)))^2\lambda\Big(K_M k_{on}^2(k_{on}+K_M k_{on}n_X+K_M\lambda)+\\
&K_A k_{on}\Big(k_{on}^2\Big(1+3K_M^2 n_X(n_L+n_X)+2K_M(n_L+2n_X)\Big)+K_M k_{on}(3+2K_M n_L+5K_M n_X)\lambda+2K_M^2\lambda^2\Big)+\\
&K_A^2\Big(k_{on}^3 n_X\Big(1+3K_M(n_L+n_X)+2K_M^2(n_L+n_X)^2\Big)+k_{on}^2\\
&\quad\Big(1+K_M(2n_L+5n_X)+K_M^2\Big(n_L^2+6n_L n_X+5n_X^2\Big)\Big)\lambda+2K_M k_{on}(1+K_M(n_L+2n_X))\lambda^2+K_M^2\lambda^3\Big)\Big)\Big)
\end{aligned}$$

```
Print["σ^2_Num=", var /. substex // Numerator // Expand, "\n"]
Print["σ^2_Den=", var /. substex // Denominator // Expand, "\n"]
```

$$
\begin{aligned}
\sigma_{Num}^2 = {} & g_{on} K_A k_{on}^3 + g_{on} K_M k_{on}^3 + 2 g_{on} K_A K_M k_{on}^3 n_L + g_{on}^2 K_A K_M^2 k_{on}^2 n_X + g_{on}^2 K_M^3 k_{on}^2 n_X + 3 g_{on} K_A^2 k_{on}^3 n_X + \\
& 7 g_{on} K_A K_M k_{on}^3 n_X + 2 g_{on} K_M^2 k_{on}^3 n_X + g_{on}^2 K_A^2 K_M^2 k_{on}^2 n_L n_X + 3 g_{on}^2 K_A K_M^3 k_{on}^2 n_L n_X + \\
& 8 g_{on} K_A^2 K_M k_{on}^3 n_L n_X + 6 g_{on} K_A K_M^2 k_{on}^3 n_L n_X + g_{on}^2 K_A^3 K_M^2 k_{on}^2 n_L^2 n_X + 3 g_{on}^2 K_A^2 K_M^3 k_{on}^2 n_L^2 n_X + \\
& 4 g_{on} K_A^2 K_M^2 k_{on}^3 n_L^2 n_X + g_{on}^2 K_A^3 K_M^3 k_{on}^2 n_L^3 n_X + 3 g_{on}^2 K_A^2 K_M^2 k_{on}^2 n_X^2 + 5 g_{on}^2 K_A K_M^3 k_{on}^2 n_X^2 + \\
& 3 g_{on} K_A^3 k_{on}^3 n_X^2 + 15 g_{on} K_A^2 K_M k_{on}^3 n_X^2 + 11 g_{on} K_A K_M^2 k_{on}^3 n_X^2 + g_{on} K_M^3 k_{on}^3 n_X^2 + \\
& 2 g_{on}^2 K_A^3 K_M^2 k_{on}^2 n_L n_X^2 + 10 g_{on}^2 K_A^2 K_M^3 k_{on}^2 n_L n_X^2 + 10 g_{on} K_A^3 K_M k_{on}^3 n_L n_X^2 + 22 g_{on} K_A^2 K_M^2 k_{on}^3 n_L n_X^2 + \\
& 4 g_{on} K_A K_M^3 k_{on}^3 n_L n_X^2 + 5 g_{on}^2 K_A^3 K_M^3 k_{on}^2 n_L^2 n_X^2 + 9 g_{on} K_A^3 K_M^2 k_{on}^3 n_L^2 n_X^2 + 5 g_{on} K_A^2 K_M^3 k_{on}^3 n_L^2 n_X^2 + \\
& 2 g_{on} K_A^3 K_M^3 k_{on}^3 n_L^3 n_X^2 + 3 g_{on}^2 K_A^3 K_M^2 k_{on}^2 n_X^3 + 9 g_{on}^2 K_A^2 K_M^3 k_{on}^2 n_X^3 + g_{on} K_A^4 k_{on}^3 n_X^3 + \\
& 13 g_{on} K_A^3 K_M k_{on}^3 n_X^3 + 21 g_{on} K_A^2 K_M^2 k_{on}^3 n_X^3 + 5 g_{on} K_A K_M^3 k_{on}^3 n_X^3 + g_{on}^2 K_A^4 K_M^2 k_{on}^2 n_L n_X^3 + \\
& 11 g_{on}^2 K_A^3 K_M^3 k_{on}^2 n_L n_X^3 + 4 g_{on} K_A^4 K_M k_{on}^3 n_L n_X^3 + 26 g_{on} K_A^3 K_M^2 k_{on}^3 n_L n_X^3 + 14 g_{on} K_A^2 K_M^3 k_{on}^3 n_L n_X^3 + \\
& 2 g_{on}^2 K_A^4 K_M^3 k_{on}^2 n_L^2 n_X^3 + 5 g_{on} K_A^4 K_M^2 k_{on}^3 n_L^2 n_X^3 + 11 g_{on} K_A^3 K_M^3 k_{on}^3 n_L^2 n_X^3 + 2 g_{on} K_A^4 K_M^3 k_{on}^3 n_L^3 n_X^3 + \\
& g_{on}^2 K_A^4 K_M^2 k_{on}^2 n_X^4 + 7 g_{on}^2 K_A^3 K_M^3 k_{on}^2 n_X^4 + 4 g_{on} K_A^4 K_M k_{on}^3 n_X^4 + 17 g_{on} K_A^3 K_M^2 k_{on}^3 n_X^4 + \\
& 9 g_{on} K_A^2 K_M^3 k_{on}^3 n_X^4 + 4 g_{on}^2 K_A^4 K_M^3 k_{on}^2 n_L n_X^4 + 10 g_{on} K_A^4 K_M^2 k_{on}^3 n_L n_X^4 + 16 g_{on} K_A^3 K_M^3 k_{on}^3 n_L n_X^4 + \\
& 6 g_{on} K_A^4 K_M^3 k_{on}^3 n_L^2 n_X^4 + 2 g_{on}^2 K_A^4 K_M^3 k_{on}^2 n_X^5 + 5 g_{on} K_A^4 K_M^2 k_{on}^3 n_X^5 + 7 g_{on} K_A^3 K_M^3 k_{on}^3 n_X^5 + \\
& 6 g_{on} K_A^4 K_M^3 k_{on}^3 n_L n_X^5 + 2 g_{on} K_A^4 K_M^3 k_{on}^3 n_X^6 + g_{on} K_A^2 k_{on}^2 \lambda + 3 g_{on} K_A K_M k_{on}^2 \lambda + g_{on} K_M^2 k_{on}^2 \lambda + \\
& 2 g_{on} K_A^2 K_M k_{on}^2 n_L \lambda + 2 g_{on} K_A K_M^2 k_{on}^2 n_L \lambda + g_{on} K_A^2 K_M^2 k_{on}^2 n_L^2 \lambda + g_{on}^2 K_A^2 K_M^2 k_{on} n_X \lambda + \\
& 2 g_{on}^2 K_A K_M^3 k_{on} n_X \lambda + 2 g_{on} K_A^3 k_{on}^2 n_X \lambda + 12 g_{on} K_A^2 K_M k_{on}^2 n_X \lambda + 10 g_{on} K_A K_M^2 k_{on}^2 n_X \lambda + \\
& g_{on} K_M^3 k_{on}^2 n_X \lambda + g_{on}^2 K_A^3 K_M^2 k_{on} n_L n_X \lambda + 4 g_{on}^2 K_A^2 K_M^3 k_{on} n_L n_X \lambda + 5 g_{on} K_A^3 K_M k_{on}^2 n_L n_X \lambda + \\
& 15 g_{on} K_A^2 K_M^2 k_{on}^2 n_L n_X \lambda + 3 g_{on} K_A K_M^3 k_{on}^2 n_L n_X \lambda + 2 g_{on}^2 K_A^3 K_M^3 k_{on} n_L^2 n_X \lambda + \\
& 4 g_{on} K_A^3 K_M^2 k_{on}^2 n_L^2 n_X \lambda + 3 g_{on} K_A^2 K_M^3 k_{on}^2 n_L^2 n_X \lambda + g_{on} K_A^3 K_M^3 k_{on}^2 n_L^3 n_X \lambda + 2 g_{on}^2 K_A^3 K_M^2 k_{on} n_X^2 \lambda + \\
& 7 g_{on}^2 K_A^2 K_M^3 k_{on} n_X^2 \lambda + g_{on} K_A^4 k_{on}^2 n_X^2 \lambda + 15 g_{on} K_A^3 K_M k_{on}^2 n_X^2 \lambda + 27 g_{on} K_A^2 K_M^2 k_{on}^2 n_X^2 \lambda + \\
& 7 g_{on} K_A K_M^3 k_{on}^2 n_X^2 \lambda + g_{on}^2 K_A^4 K_M^2 k_{on} n_L n_X^2 \lambda + 8 g_{on}^2 K_A^3 K_M^3 k_{on} n_L n_X^2 \lambda + 3 g_{on} K_A^4 K_M k_{on}^2 n_L n_X^2 \lambda + \\
& 26 g_{on} K_A^3 K_M^2 k_{on}^2 n_L n_X^2 \lambda + 16 g_{on} K_A^2 K_M^3 k_{on}^2 n_L n_X^2 \lambda + g_{on}^2 K_A^4 K_M^3 k_{on} n_L^2 n_X^2 \lambda + \\
& 3 g_{on} K_A^4 K_M^2 k_{on}^2 n_L^2 n_X^2 \lambda + 10 g_{on} K_A^3 K_M^3 k_{on}^2 n_L^2 n_X^2 \lambda + g_{on} K_A^4 K_M^3 k_{on}^2 n_L^3 n_X^2 \lambda + g_{on}^2 K_A^4 K_M^2 k_{on} n_X^3 \lambda + \\
& 8 g_{on}^2 K_A^3 K_M^3 k_{on} n_X^3 \lambda + 6 g_{on} K_A^4 K_M k_{on}^2 n_X^3 \lambda + 28 g_{on} K_A^3 K_M^2 k_{on}^2 n_X^3 \lambda + 16 g_{on} K_A^2 K_M^3 k_{on}^2 n_X^3 \lambda + \\
& 4 g_{on}^2 K_A^4 K_M^3 k_{on} n_L n_X^3 \lambda + 13 g_{on} K_A^4 K_M^2 k_{on}^2 n_L n_X^3 \lambda + 24 g_{on} K_A^3 K_M^3 k_{on}^2 n_L n_X^3 \lambda + \\
& 7 g_{on} K_A^4 K_M^3 k_{on}^2 n_L^2 n_X^3 \lambda + 3 g_{on}^2 K_A^4 K_M^3 k_{on} n_X^4 \lambda + 10 g_{on} K_A^4 K_M^2 k_{on}^2 n_X^4 \lambda + 15 g_{on} K_A^3 K_M^3 k_{on}^2 n_X^4 \lambda + \\
& 11 g_{on} K_A^4 K_M^3 k_{on}^2 n_L n_X^4 \lambda + 5 g_{on} K_A^4 K_M^3 k_{on}^2 n_X^5 \lambda + 2 g_{on} K_A^2 K_M k_{on} \lambda^2 + 2 g_{on} K_A K_M^2 k_{on} \lambda^2 + \\
& 2 g_{on} K_A^2 K_M^2 k_{on} n_L \lambda^2 + g_{on}^2 K_A^2 K_M^3 n_X \lambda^2 + 4 g_{on} K_A^3 K_M k_{on} n_X \lambda^2 + 10 g_{on} K_A^2 K_M^2 k_{on} n_X \lambda^2 + \\
& 2 g_{on} K_A K_M^3 k_{on} n_X \lambda^2 + g_{on}^2 K_A^3 K_M^3 n_L n_X \lambda^2 + 6 g_{on} K_A^3 K_M^2 k_{on} n_L n_X \lambda^2 + 4 g_{on} K_A^2 K_M^3 k_{on} n_L n_X \lambda^2 + \\
& 2 g_{on} K_A^3 K_M^3 k_{on} n_L^2 n_X \lambda^2 + 2 g_{on}^2 K_A^3 K_M^3 n_X^2 \lambda^2 + 2 g_{on} K_A^4 K_M k_{on} n_X^2 \lambda^2 + 14 g_{on} K_A^3 K_M^2 k_{on} n_X^2 \lambda^2 + \\
& 8 g_{on} K_A^2 K_M^3 k_{on} n_X^2 \lambda^2 + g_{on}^2 K_A^4 K_M^3 n_L n_X^2 \lambda^2 + 4 g_{on} K_A^4 K_M^2 k_{on} n_L n_X^2 \lambda^2 + 10 g_{on} K_A^3 K_M^3 k_{on} n_L n_X^2 \lambda^2 + \\
& 2 g_{on} K_A^4 K_M^3 k_{on} n_L^2 n_X^2 \lambda^2 + g_{on}^2 K_A^4 K_M^3 n_X^3 \lambda^2 + 6 g_{on} K_A^4 K_M^2 k_{on} n_X^3 \lambda^2 + 10 g_{on} K_A^3 K_M^3 k_{on} n_X^3 \lambda^2 + \\
& 6 g_{on} K_A^4 K_M^3 k_{on} n_L n_X^3 \lambda^2 + 4 g_{on} K_A^4 K_M^3 k_{on} n_X^4 \lambda^2 + g_{on} K_A^2 K_M^2 \lambda^3 + 2 g_{on} K_A^3 K_M^2 n_X \lambda^3 + g_{on} K_A^2 K_M^3 n_X \lambda^3 + \\
& g_{on} K_A^3 K_M^3 n_L n_X \lambda^3 + g_{on} K_A^4 K_M^2 n_X^2 \lambda^3 + 2 g_{on} K_A^3 K_M^3 n_X^2 \lambda^3 + g_{on} K_A^4 K_M^3 n_L n_X^2 \lambda^3 + g_{on} K_A^4 K_M^3 n_X^3 \lambda^3
\end{aligned}
$$

$$\sigma^2_{Den} = K_A k_{on}^3 \lambda + K_M k_{on}^3 \lambda + 2 K_A K_M k_{on}^3 n_L \lambda + 3 K_A^2 k_{on}^3 n_X \lambda + 8 K_A K_M k_{on}^3 n_X \lambda + 3 K_M^2 k_{on}^3 n_X \lambda +$$
$$9 K_A^2 K_M k_{on}^3 n_L n_X \lambda + 9 K_A K_M^2 k_{on}^3 n_L n_X \lambda + 6 K_A^2 K_M^2 k_{on}^3 n_L^2 n_X \lambda + 3 K_A^3 k_{on}^3 n_X^2 \lambda + 18 K_A^2 K_M k_{on}^3 n_X^2 \lambda +$$
$$18 K_A K_M^2 k_{on}^3 n_X^2 \lambda + 3 K_M^3 k_{on}^3 n_X^2 \lambda + 12 K_A^3 K_M k_{on}^3 n_L n_X^2 \lambda + 36 K_A^2 K_M^2 k_{on}^3 n_L n_X^2 \lambda +$$
$$12 K_A K_M^3 k_{on}^3 n_L n_X^2 \lambda + 15 K_A^3 K_M^2 k_{on}^3 n_L^2 n_X^2 \lambda + 15 K_A^2 K_M^3 k_{on}^3 n_L^2 n_X^2 \lambda + 6 K_A^3 K_M^3 k_{on}^3 n_L^3 n_X^2 \lambda +$$
$$K_A^4 k_{on}^3 n_X^3 \lambda + 16 K_A^3 K_M k_{on}^3 n_X^3 \lambda + 36 K_A^2 K_M^2 k_{on}^3 n_X^3 \lambda + 16 K_A K_M^3 k_{on}^3 n_X^3 \lambda + K_M^4 k_{on}^3 n_X^3 \lambda +$$
$$5 K_A^4 K_M k_{on}^3 n_L n_X^3 \lambda + 45 K_A^3 K_M^2 k_{on}^3 n_L n_X^3 \lambda + 45 K_A^2 K_M^3 k_{on}^3 n_L n_X^3 \lambda + 5 K_A K_M^4 k_{on}^3 n_L n_X^3 \lambda +$$
$$9 K_A^4 K_M^2 k_{on}^3 n_L^2 n_X^3 \lambda + 36 K_A^3 K_M^3 k_{on}^3 n_L^2 n_X^3 \lambda + 9 K_A^2 K_M^4 k_{on}^3 n_L^2 n_X^3 \lambda + 7 K_A^4 K_M^3 k_{on}^3 n_L^3 n_X^3 \lambda +$$
$$7 K_A^3 K_M^4 k_{on}^3 n_L^3 n_X^3 \lambda + 2 K_A^4 K_M^4 k_{on}^3 n_L^4 n_X^3 \lambda + 5 K_A^4 K_M k_{on}^3 n_X^4 \lambda + 30 K_A^3 K_M^2 k_{on}^3 n_X^4 \lambda +$$
$$30 K_A^2 K_M^3 k_{on}^3 n_X^4 \lambda + 5 K_A K_M^4 k_{on}^3 n_X^4 \lambda + 18 K_A^4 K_M^2 k_{on}^3 n_L n_X^4 \lambda + 54 K_A^3 K_M^3 k_{on}^3 n_L n_X^4 \lambda +$$
$$18 K_A^2 K_M^4 k_{on}^3 n_L n_X^4 \lambda + 21 K_A^4 K_M^3 k_{on}^3 n_L^2 n_X^4 \lambda + 21 K_A^3 K_M^4 k_{on}^3 n_L^2 n_X^4 \lambda + 8 K_A^4 K_M^4 k_{on}^3 n_L^3 n_X^4 \lambda +$$
$$9 K_A^4 K_M^2 k_{on}^3 n_X^5 \lambda + 24 K_A^3 K_M^3 k_{on}^3 n_X^5 \lambda + 9 K_A^2 K_M^4 k_{on}^3 n_X^5 \lambda + 21 K_A^4 K_M^3 k_{on}^3 n_L n_X^5 \lambda +$$
$$21 K_A^3 K_M^4 k_{on}^3 n_L n_X^5 \lambda + 12 K_A^4 K_M^4 k_{on}^3 n_L^2 n_X^5 \lambda + 7 K_A^4 K_M^3 k_{on}^3 n_X^6 \lambda + 7 K_A^3 K_M^4 k_{on}^3 n_X^6 \lambda +$$
$$8 K_A^4 K_M^4 k_{on}^3 n_L n_X^6 \lambda + 2 K_A^4 K_M^4 k_{on}^3 n_X^7 \lambda + K_A^2 k_{on}^2 \lambda^2 + 3 K_A K_M k_{on}^2 \lambda^2 + K_M^2 k_{on}^2 \lambda^2 +$$
$$2 K_A^2 K_M k_{on}^2 n_L \lambda^2 + 2 K_A K_M^2 k_{on}^2 n_L \lambda^2 + K_A^2 K_M^2 k_{on}^2 n_L^2 \lambda^2 + 2 K_A^3 k_{on}^2 n_X \lambda^2 + 13 K_A^2 K_M k_{on}^2 n_X \lambda^2 +$$
$$13 K_A K_M^2 k_{on}^2 n_X \lambda^2 + 2 K_M^3 k_{on}^2 n_X \lambda^2 + 6 K_A^3 K_M k_{on}^2 n_L n_X \lambda^2 + 20 K_A^2 K_M^2 k_{on}^2 n_L n_X \lambda^2 +$$
$$6 K_A K_M^3 k_{on}^2 n_L n_X \lambda^2 + 6 K_A^3 K_M^2 k_{on}^2 n_L^2 n_X \lambda^2 + 6 K_A^2 K_M^3 k_{on}^2 n_L^2 n_X \lambda^2 + 2 K_A^3 K_M^3 k_{on}^2 n_L^3 n_X \lambda^2 +$$
$$K_A^4 k_{on}^2 n_X^2 \lambda^2 + 17 K_A^3 K_M k_{on}^2 n_X^2 \lambda^2 + 39 K_A^2 K_M^2 k_{on}^2 n_X^2 \lambda^2 + 17 K_A K_M^3 k_{on}^2 n_X^2 \lambda^2 + K_M^4 k_{on}^2 n_X^2 \lambda^2 +$$
$$4 K_A^4 K_M k_{on}^2 n_L n_X^2 \lambda^2 + 40 K_A^3 K_M^2 k_{on}^2 n_L n_X^2 \lambda^2 + 40 K_A^2 K_M^3 k_{on}^2 n_L n_X^2 \lambda^2 + 4 K_A K_M^4 k_{on}^2 n_L n_X^2 \lambda^2 +$$
$$6 K_A^4 K_M^2 k_{on}^2 n_L^2 n_X^2 \lambda^2 + 27 K_A^3 K_M^3 k_{on}^2 n_L^2 n_X^2 \lambda^2 + 6 K_A^2 K_M^4 k_{on}^2 n_L^2 n_X^2 \lambda^2 + 4 K_A^4 K_M^3 k_{on}^2 n_L^3 n_X^2 \lambda^2 +$$
$$4 K_A^3 K_M^4 k_{on}^2 n_L^3 n_X^2 \lambda^2 + K_A^4 K_M^4 k_{on}^2 n_L^4 n_X^2 \lambda^2 + 7 K_A^4 K_M k_{on}^2 n_X^3 \lambda^2 + 43 K_A^3 K_M^2 k_{on}^2 n_X^3 \lambda^2 +$$
$$43 K_A^2 K_M^3 k_{on}^2 n_X^3 \lambda^2 + 7 K_A K_M^4 k_{on}^2 n_X^3 \lambda^2 + 22 K_A^4 K_M^2 k_{on}^2 n_L n_X^3 \lambda^2 + 68 K_A^3 K_M^3 k_{on}^2 n_L n_X^3 \lambda^2 +$$
$$22 K_A^2 K_M^4 k_{on}^2 n_L n_X^3 \lambda^2 + 23 K_A^4 K_M^3 k_{on}^2 n_L^2 n_X^3 \lambda^2 + 23 K_A^3 K_M^4 k_{on}^2 n_L^2 n_X^3 \lambda^2 + 8 K_A^4 K_M^4 k_{on}^2 n_L^3 n_X^3 \lambda^2 +$$
$$16 K_A^4 K_M^2 k_{on}^2 n_X^4 \lambda^2 + 43 K_A^3 K_M^3 k_{on}^2 n_X^4 \lambda^2 + 16 K_A^2 K_M^4 k_{on}^2 n_X^4 \lambda^2 + 34 K_A^4 K_M^3 k_{on}^2 n_L n_X^4 \lambda^2 +$$
$$34 K_A^3 K_M^4 k_{on}^2 n_L n_X^4 \lambda^2 + 18 K_A^4 K_M^4 k_{on}^2 n_L^2 n_X^4 \lambda^2 + 15 K_A^4 K_M^3 k_{on}^2 n_X^5 \lambda^2 + 15 K_A^3 K_M^4 k_{on}^2 n_X^5 \lambda^2 +$$
$$16 K_A^4 K_M^4 k_{on}^2 n_L n_X^5 \lambda^2 + 5 K_A^4 K_M^4 k_{on}^2 n_X^6 \lambda^2 + 2 K_A^2 K_M k_{on} \lambda^3 + 2 K_A K_M^2 k_{on} \lambda^3 + 2 K_A^2 K_M^2 k_{on} n_L \lambda^3 +$$
$$4 K_A^3 K_M k_{on} n_X \lambda^3 + 12 K_A^2 K_M^2 k_{on} n_X \lambda^3 + 4 K_A K_M^3 k_{on} n_X \lambda^3 + 8 K_A^3 K_M^2 k_{on} n_L n_X \lambda^3 + 8 K_A^2 K_M^3 k_{on} n_L n_X \lambda^3 +$$
$$4 K_A^3 K_M^3 k_{on} n_L^2 n_X \lambda^3 + 2 K_A^4 K_M k_{on} n_X^2 \lambda^3 + 18 K_A^3 K_M^2 k_{on} n_X^2 \lambda^3 + 18 K_A^2 K_M^3 k_{on} n_X^2 \lambda^3 +$$
$$2 K_A K_M^4 k_{on} n_X^2 \lambda^3 + 6 K_A^4 K_M^2 k_{on} n_L n_X^2 \lambda^3 + 24 K_A^3 K_M^3 k_{on} n_L n_X^2 \lambda^3 + 6 K_A^2 K_M^4 k_{on} n_L n_X^2 \lambda^3 +$$
$$6 K_A^4 K_M^3 k_{on} n_L^2 n_X^2 \lambda^3 + 6 K_A^3 K_M^4 k_{on} n_L^2 n_X^2 \lambda^3 + 2 K_A^4 K_M^4 k_{on} n_L^3 n_X^2 \lambda^3 + 8 K_A^4 K_M^2 k_{on} n_X^3 \lambda^3 +$$
$$24 K_A^3 K_M^3 k_{on} n_X^3 \lambda^3 + 8 K_A^2 K_M^4 k_{on} n_X^3 \lambda^3 + 16 K_A^4 K_M^3 k_{on} n_L n_X^3 \lambda^3 + 16 K_A^3 K_M^4 k_{on} n_L n_X^3 \lambda^3 +$$
$$8 K_A^4 K_M^4 k_{on} n_L^2 n_X^3 \lambda^3 + 10 K_A^4 K_M^3 k_{on} n_X^4 \lambda^3 + 10 K_A^3 K_M^4 k_{on} n_X^4 \lambda^3 + 10 K_A^4 K_M^4 k_{on} n_L n_X^4 \lambda^3 +$$
$$4 K_A^4 K_M^4 k_{on} n_X^5 \lambda^3 + K_A^2 K_M^2 \lambda^4 + 2 K_A^3 K_M^2 n_X \lambda^4 + 2 K_A^2 K_M^3 n_X \lambda^4 + 2 K_A^3 K_M^3 n_L n_X \lambda^4 +$$
$$K_A^4 K_M^2 n_X^2 \lambda^4 + 4 K_A^3 K_M^3 n_X^2 \lambda^4 + K_A^2 K_M^4 n_X^2 \lambda^4 + 2 K_A^4 K_M^3 n_L n_X^2 \lambda^4 + 2 K_A^3 K_M^4 n_L n_X^2 \lambda^4 +$$
$$K_A^4 K_M^4 n_L^2 n_X^2 \lambda^4 + 2 K_A^4 K_M^3 n_X^3 \lambda^4 + 2 K_A^3 K_M^4 n_X^3 \lambda^4 + 2 K_A^4 K_M^4 n_L n_X^3 \lambda^4 + K_A^4 K_M^4 n_X^4 \lambda^4$$

```
m = mvs // Simplify;
f = fano // Simplify;
c = CV2 // Simplify;
```

```
LogLogPlot[Evaluate[{m /. nl → 0, m} /. {n → 1 / xx} /. parex /. pars],
 {xx, 0.1, 10^4}, AxesLabel → {"n_WT/n_X", "μ"}]
LogLogPlot[Evaluate[{f /. nl → 0, f} /. {n → 1 / xx} /. parex /. pars],
 {xx, 0.1, 10^4}, AxesLabel → {"n_WT/n_X", "σ^2/μ"}]
LogLogPlot[Evaluate[{√c /. nl → 0, √c} /. {n → 1 / xx} /. parex /. pars],
 {xx, 0.1, 10^4}, AxesLabel → {"n_WT/n_X", "σ/μ"}]
```

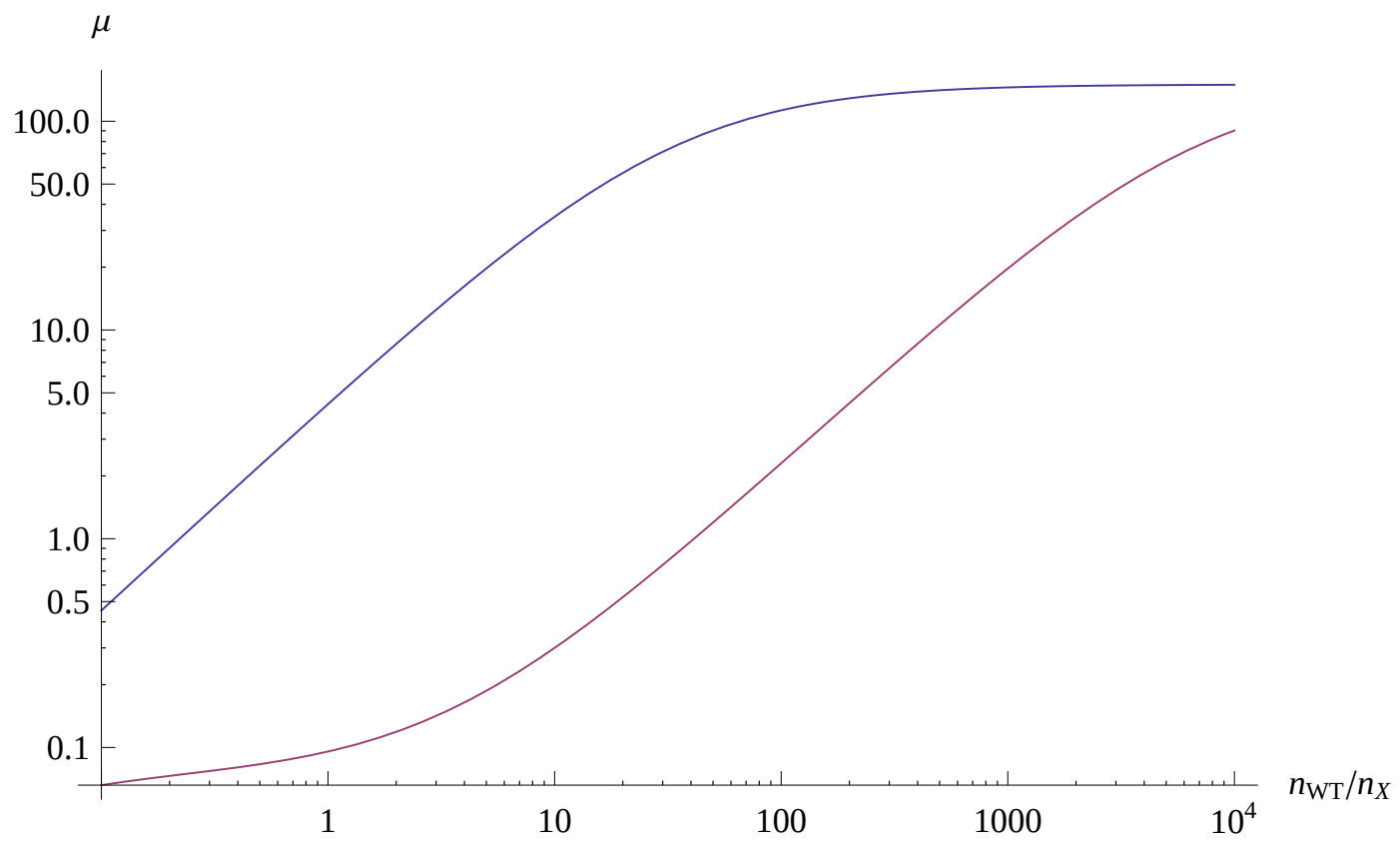

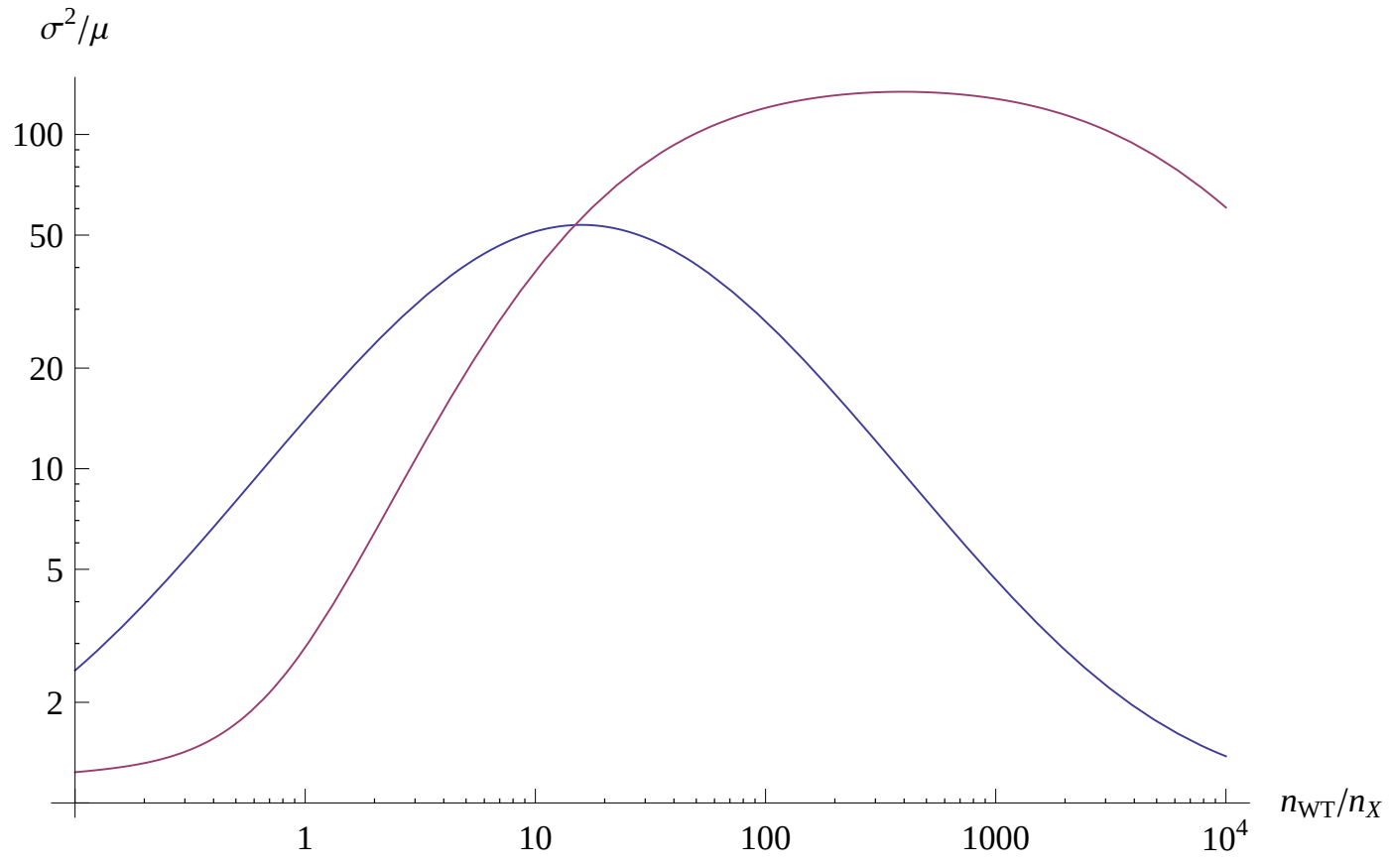

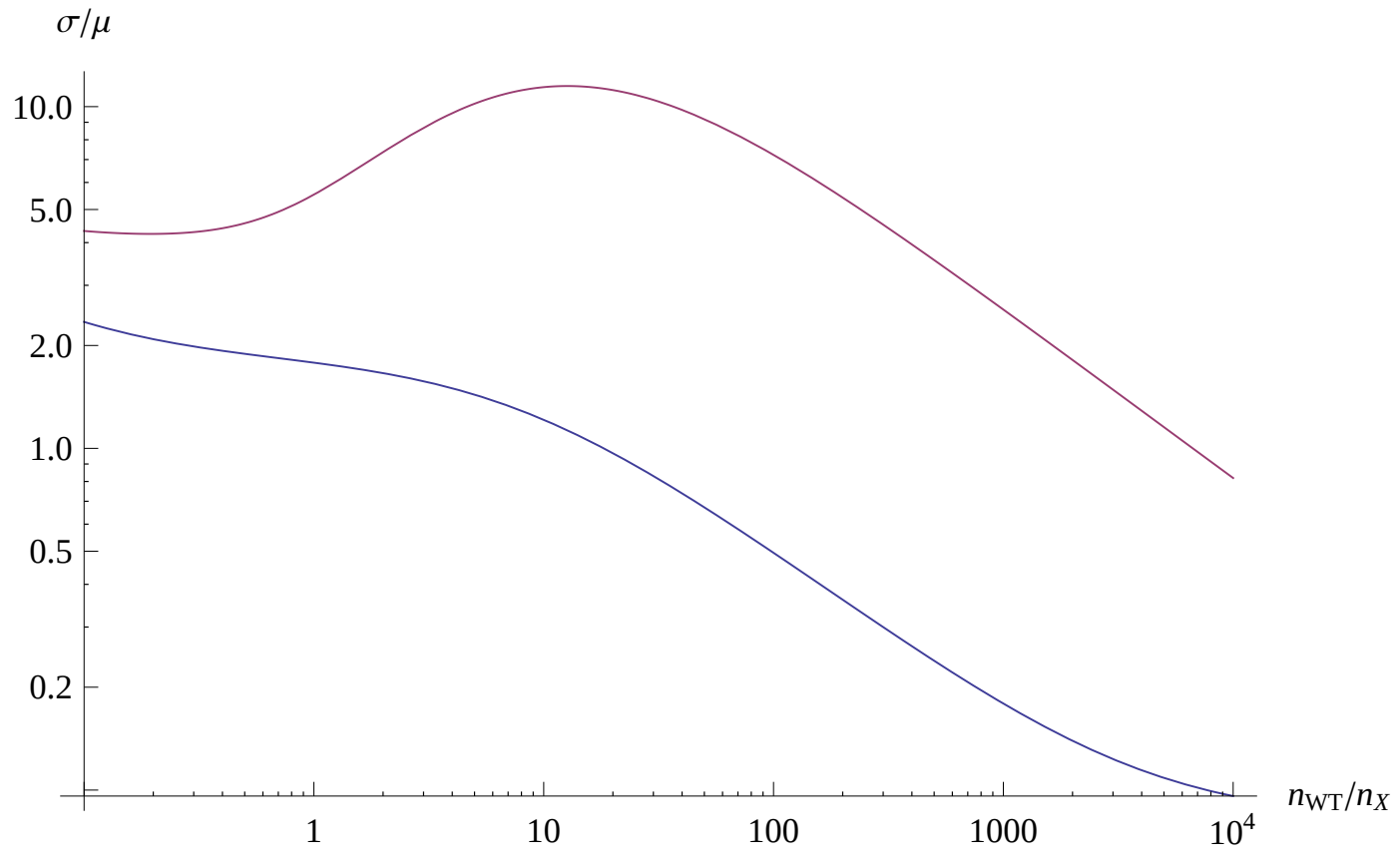